\documentclass[prb,reprint,longbibliography,nofootinbib]{revtex4-2} 

\usepackage{amsmath}
\usepackage{amsfonts}
\usepackage{graphicx}
\usepackage[colorlinks, allcolors=blue]{hyperref}
\usepackage{trfsigns}
\usepackage{siunitx}

\begin{document}
\title{Roller coaster dynamics -- from point particles to a continuum model
  using Lagrange density}

\author{Michael Kaschke}
\email{michael.kaschke@kit.edu}
\affiliation{Department Electrical Engineering and Information Technology,
  Karlsruhe Institute of Technology, 76131 Karlsruhe, Germany}
\author{Holger Cartarius}
\email{holger.cartarius@uni-jena.de}
\affiliation{Research Group Teaching Methodology in Physics and Astronomy,
  Friedrich Schiller University Jena, 07743 Jena, Germany}

\date{\today}

\begin{abstract}
  Analyzing the motion of a roller coaster allows for an instructive
  introduction of various theoretical concepts in a concrete and enjoyable
  context. We start by modeling the roller coaster train as a point 
  particle. We then develop more realistic models for the  train and finally
  we show how  to introduce a continuum limit in a simple way. These studies
  instructively illustrate the relationships between  different formalisms
  (Newtonian mechanics, Lagrangian mechanics of the first and second kind, as
  well as continuous Lagrangian mechanics using a Lagrangian density). We
  derive the equations of motion in all considered models and calculate the
  forces acting on the track and on the passengers. Numerical results are
  also provided and discussed.
\end{abstract}

\maketitle

\section{Introduction}

The physics of roller coasters is quite rich \cite{Pendrill2021,Walker1983}
and catches the interest of most students. Not surprisingly there are already
a large number of publications on the subject, using a variety of different
methodologies. Students can study roller coaster motion using video recordings
\cite{Mamola2021} or experimentally with portable sensors
\cite{Speers1991,Monteiro2022,Pendrill2023}. 

The relationship between rider experience, wagon choice, and track design can
be analyzed with the tools of classical mechanics \cite{Alberghi2007,
  Pendrill2013,Pendrill2020,Nordmark2010,Weisenberger2013}.  Entertaining
questions can be addressed, such as ``What is the optimal wagon choice if you
are interested in experiencing the strongest forces?'' \cite{Alberghi2007,
  Pendrill2013}. Roller coasters connect core mechanics concepts to an
accessible and engaging context, which is appealing for teaching
\cite{Mejia2020,Souza2017,Cross2023,Mueller2010}. 

In this work, we discuss roller coaster dynamics using various
theoretical concepts accessible to undergraduate students.  Forces
acting on the passengers can be understood with a Newtonian mechanics
point-particle description. Track constraints can be analyzed with
Lagrangian mechanics of the first kind, providing simple and
instructive access to these forces. To discuss whether a passenger in
the front, middle, or back of the train feels the strongest forces,
the train can be modeled as a line of fixed length and again analyzed
using Lagrangian mechanics. As far as we know, the fact that roller
coaster trains posses a finite (but low) elasticity has not been
addressed in the literature. Springs between train wagons can be
included to model this effect; this can be an instructive exercise in
the study of elasticity, even though it is unlikely to be a
significant effect in real roller coasters.  Finally, the roller
coaster can be extended to a model of an elastic continuous body by
starting with Lagrangian mechanics of the second kind and taking the
continuum limit; this is an ideal way to introduce students to the
concept of Lagrangian density.

The path we tread provides students an opportunity to learn about
roller coaster physics, while also allowing them to discover the
relationship between different formalisms (Newtonian mechanics,
Lagrangian mechanics of the first and second kind, as well as
continuous mechanics using a Lagrangian density) and how they build on
each other. While the equations of motions allow for a qualitative
discussion of the roller coaster's behavior, sometimes they must be
solved numerically, allowing for the introduction of numerical
exercises using, for example, MATLAB or Python.

To gain first insights into the roller coaster dynamics we start with
a simple point-particle model in Sec.~\ref{sec:point_particle} and
derive all forces relevant for our discussion. A Gaussian track is
used to illustrate the findings with numerical solutions. The
presented method can easily be applied to other more realistic track
shapes. Section \ref{sec:fixed_length} introduces a model improvement
to a train of fixed-length. In Sec.~\ref{sec:wagons_springs} we allow
the train to stretch by treating it as a chain of point-mass wagons
connected via springs. The influence of the spring's harmonic
potential on the motions of the individual wagons is shown. The
continuum limit for a train with nonvanishing elasticity is finally
performed in Sec.~\ref{sec:elastic}. This model and its numerical
solutions are compared to the previous descriptions
developed. Conclusions are drawn in Sec.~\ref{sec:conclusion}.

\section{Roller coaster as point particle}
\label{sec:point_particle}

First, we treat the roller coaster as a point particle.  Since we are
interested in the forces experienced during the ride, we use Lagrangian
mechanics of the first kind. Our roller coaster moves in the $x$-$z$ plane and
we assume that the track can be described by a function $z=h(x)$, which
provides a constraint.\footnote{In this section we use a simple track given
  by a function $z(x)$ to keep the formalism simple. Of course, roller coaster
  tracks usually contain segments of more complicated shapes such as loops,
  which require a parameterized curve. This can also be modeled by writing
  $x$ and $z$ in terms of a parameter $u$. For details and an example we refer
  the reader to Sec.~S-III~A of the supplementary material.}
The constraining equation is given by
\begin{align}
  f = z-h(x) = 0 \; ,
  \label{eq:constraint_point_particle}
\end{align}
which leads to the equations of motion
\begin{subequations}
  \begin{align}
    m\,\ddot{x} &= m \lambda \,\frac{\partial f}{\partial x}
                  = - m \lambda\,\frac{\partial h}{\partial x}
                  = - m \lambda\,h'(x) \;, 
                  \label{eq:eqm_point_a} \\
    m\,\ddot{z} &= -m\,g + m \lambda\,\frac{\partial f}{\partial z}
                  = -m\,g + m \lambda \;,
                  \label{eq:eqm_point_b}
  \end{align}
\end{subequations}
where $g$ is the gravitational acceleration, and the traditional Lagrange
multiplier $\lambda$ has been replaced by $m\,\lambda$ to simplify some of
the following equations, the dot denotes a time derivative, and a prime
denotes derivatives with respect to $x$.

By taking time derivatives of
\eqref{eq:constraint_point_particle}, we have
\begin{subequations}
  \begin{align}
    0 = \dot{f}
    &= \dot{z} - h'(x)\,\dot{x} \; ,
      \label{eq:point_derivative_constraint_a} \\
    0 = \ddot{f}
    &= \ddot{z} - h''(x)\,\dot{x}^2
      - h'(x)\,\ddot{x} \; .
      \label{eq:point_derivative_constraint_b}
  \end{align}
\end{subequations}
Using this, along with the equations of motion \eqref{eq:eqm_point_a} and
\eqref{eq:eqm_point_b} the Lagrangian multiplier can be determined to
be
\begin{align}
  \lambda (x,  \dot{x}) 
  = \frac{g + h''\,\dot{x}^2}{1+ h'^2} \; ,
  \label{eq:eqm_point_x} 
\end{align}
which leads to the equations of motion in the final form
\begin{subequations}
  \begin{align}
    \ddot{x} &= - \frac{g + h''\,\dot{x}^2}{1+h'^2} \, h'
               \label{eq:eqs_motion_pointx}\,, \\
    \ddot{z} &= - g + \frac{g + h''\,\dot{x}^2}{1+ h'^2}
               \label{eq:eqs_motion_pointz} \; . 
  \end{align}
\end{subequations}
  
The ordinary differential equations \eqref{eq:eqs_motion_pointx},
\eqref{eq:eqs_motion_pointz} can be integrated numerically.  To obtain the
normal force the track exerts on the train, we first write the equation of
motion in terms of the gravitational force $\vec{F}_\text{G}$ and the normal
force: $\vec{F}_G + \vec{F}_N = m \ddot{\vec{r}}$ and then rearrange:
\begin{align}
  \vec{F}_{N}
  &= m \begin{pmatrix} \ddot{x} \\ \ddot{z} + g \end{pmatrix} \notag \\
  &= \frac{m}{1 + (h')^2}
    \begin{pmatrix}
      -\left (h''\,\dot{x}^2 + g\right )\,h' \\
      h''\,\dot{x}^2 + g
    \end{pmatrix} \notag \\
  &= \frac{m}{1 + \tan^2\theta}
    \begin{pmatrix}
      -\left ( h''\,\dot{x}^2 + g\right )\,\tan\theta \\
      h''\,\dot{x}^2 + g 
    \end{pmatrix} \;. 
\end{align} 
In the last step we introduced the angle $\theta$ which the train's trajectory
makes with the $x$-axis
\begin{equation}
  \tan\theta = h' = \frac{\partial h}{\partial x} = \frac{\partial z}
  {\partial x}.
  \label{eq:tangent_angle}
\end{equation}

Our formalism allows for realistic track shapes, but as a simple example which
illustrates the main effects, we consider a Gaussian track,
\begin{equation}
  h(x) = H\,\e^{-a^2\,x^2} \; . \label{eq:gaussian}
\end{equation}
By energy conservation, in order to make it over the top, the train's initial
velocity must satisfy:
\begin{align}
  \dot{x}(0) &=\dot{x}_0 \geq \sqrt{\frac{2\,g\,H}{1 + h'(x_0)^2}}
  = \dot{x}_{0,\mathrm{min}}\; , \\
  \dot{z}(0)&=\dot{z}_0 = h'(x_0) \,\dot{x}_0 \;.
     \label{eq:dotxmin}
\end{align}

\begin{figure}[tbp]
  \includegraphics[width=\columnwidth]{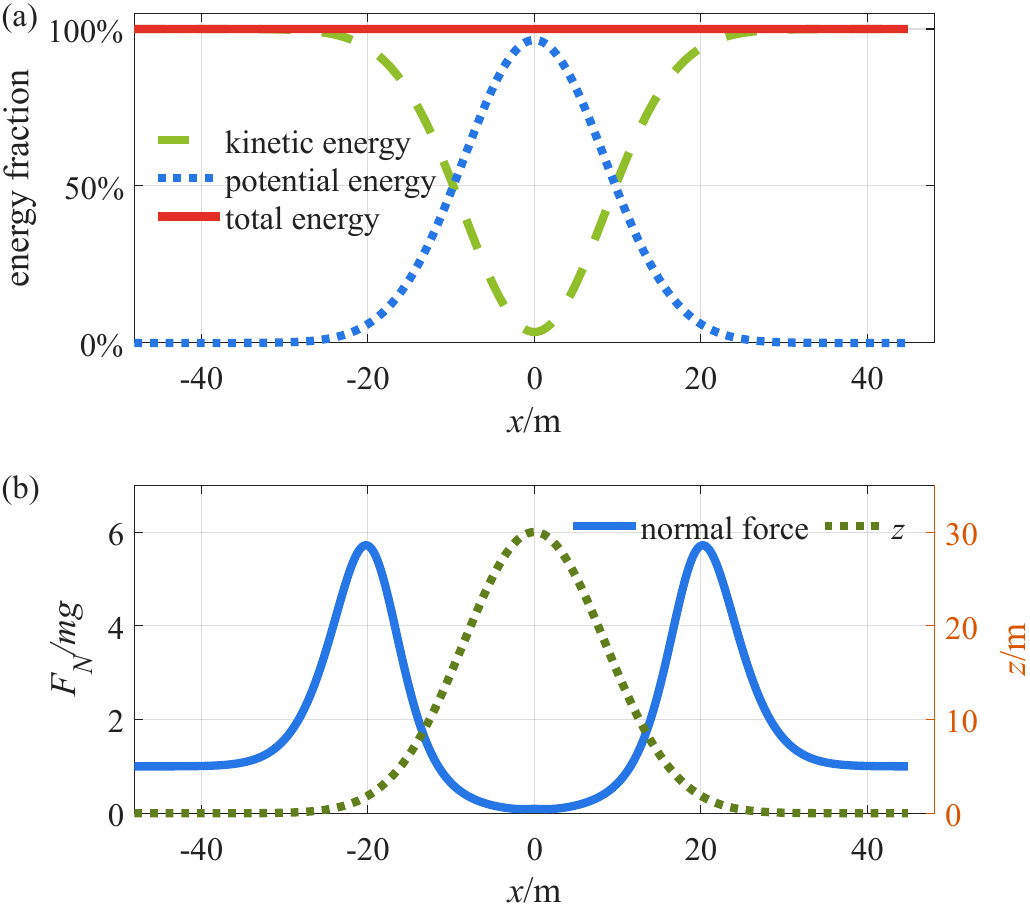}
  \caption{(a) Energy distribution and (b) normal force on a roller
    coaster train modeled by a point mass on a Gaussian track. The dotted
    green line denotes the track shape. Parameters chosen in this example
    are a height of $H = \SI{30}{\metre}$ and a half width at half maximum
    of $\SI{10}{\metre}$, i.e., $a = \sqrt{\ln 2}/\SI{10}{\metre}$. The
    initial conditions are $x_0 = -4/a$, $z_0 = h(x_0)$, and $\dot{x}_0 =
    1.018 \times \dot{x}_{0,\mathrm{min}}$. Since the initial velocity is
    only slightly above the minimum value, the velocity at the peak of the
    Gaussian almost vanishes.}
  \label{fig:point_energy_acc}
\end{figure}

For this Gaussian track, the kinetic and potential energy of the train
and the normal force are shown in Fig.\ \ref{fig:point_energy_acc}.

\begin{figure}[tbp]
  \includegraphics[width=\columnwidth]{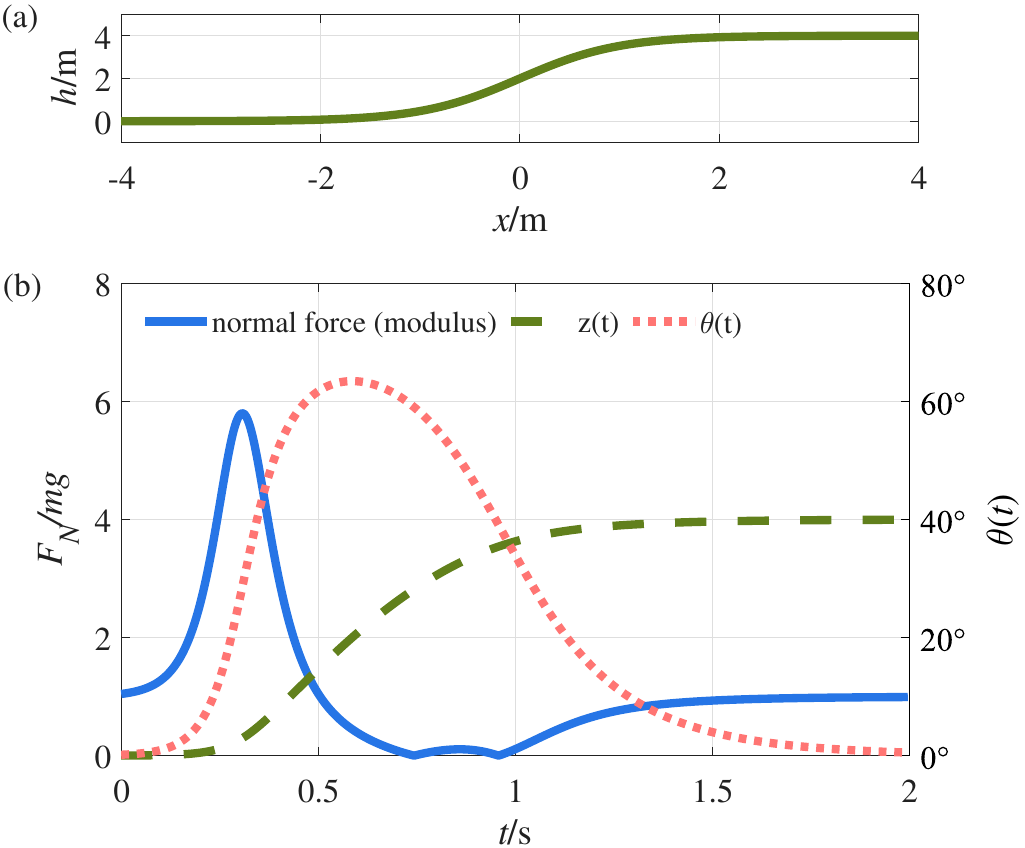}
  \caption{(a) The profile of a step-like track.  (b) The modulus of the
    normal force, track height and tangent angle for the step-like track. The
    two kinks in the modulus of the normal force result from a change of its
    direction. The normal vector is defined to point always upward away from
    the track, which is identical to the direction of the normal force at most
    points on the track. Between the two kinks, the normal force points downward
    toward the track. The initial velocity is $9.0\,\mathrm{m/s}$.}
  \label{point_normal_forces}
\end{figure}

In Fig.~\ref{point_normal_forces}
we {plot results for} a step-like track; this illustrates an
interesting situation where the direction of the normal force changes.
{When the normal force points perpendicularly \emph{toward} the track,
we call this} ``\emph{airtime}'' because if the car were not
constrained to the track, it would lose contact. During these phases
the passenger is held in the car solely by the security bar in the
wagon.  This airtime appears in between $0.5\, \mathrm{s}$ and
$1.0\,\mathrm{s}$. As soon as $\theta$ starts to decrease, i.e., the
curvature becomes negative, the normal force changes its direction
(which is identified by the kink in the modulus), and does this again
when the plateau is reached.
\section{Train model 1: Extended train of fixed length}
\label{sec:fixed_length}

The point particle model considered so far does not explain why typically the
first and last wagon on the train are the most favored by thrillseekers.  To
study this question, we introduce an extended model of a train having a fixed
length and consisting of several wagons. It is assumed that all wagons have
the same instantaneous velocity at any time. This, however, does not mean that
each wagon has the same velocity when it passes through a certain point on the
track.

\begin{figure}[tbp]
  \includegraphics[width=\columnwidth]{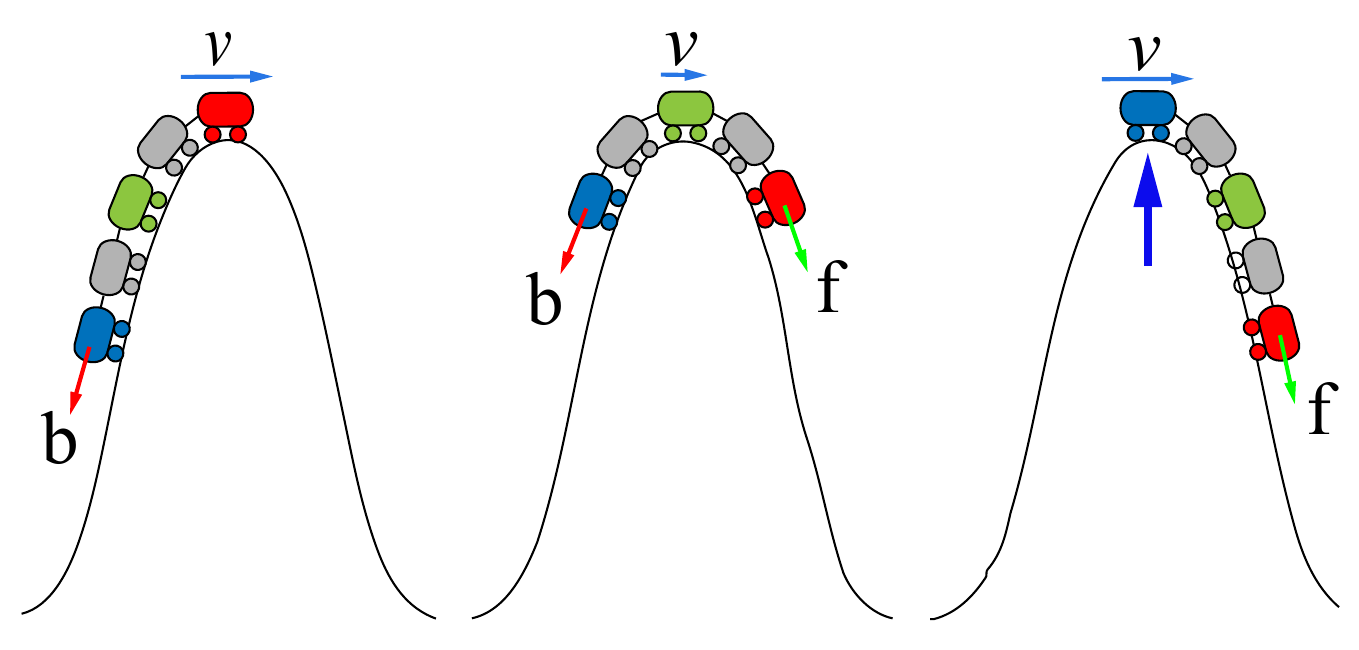}
  \caption{Roller coaster train -- ``hilltop phase'' at three
    successive times.  The thick dark blue arrow marks the position at
    which the passengers in the last wagon feel an \emph{airtime}, i.e.,
    they are pressed to the track by the security bar in the wagon and
    would lose contact with the seat without it. The lengths of the light
    blue arrows (denoted by $v$) indicate the velocity of the wagon at the
    hilltop.  Arrows at the outer wagons indicate the direction of the
    tangential component of the gravitational force on these wagons. Green
    arrows with label f represent tangential components in forward
    direction, whereas red arrows with label b show tangential forces in
    backward direction. }
  \label{fig:train_hill}
\end{figure}

Fig.~\ref{fig:train_hill}
shows a five-wagon train at three successive times. The tangential
acceleration is negative until the middle wagon is at the hilltop, and
at this point the middle wagon's speed is at a minimum. From this time on the
tangential acceleration is positive.  Consequently, the airtime felt in the
outer wagons is more intense than that felt in the middle wagon because the
outer wagons have a higher speed at the top. As we will see in the next
section, the last wagon is the fastest when elastic forces are added.

\begin{figure}[tbp]
  \includegraphics[width=\columnwidth]{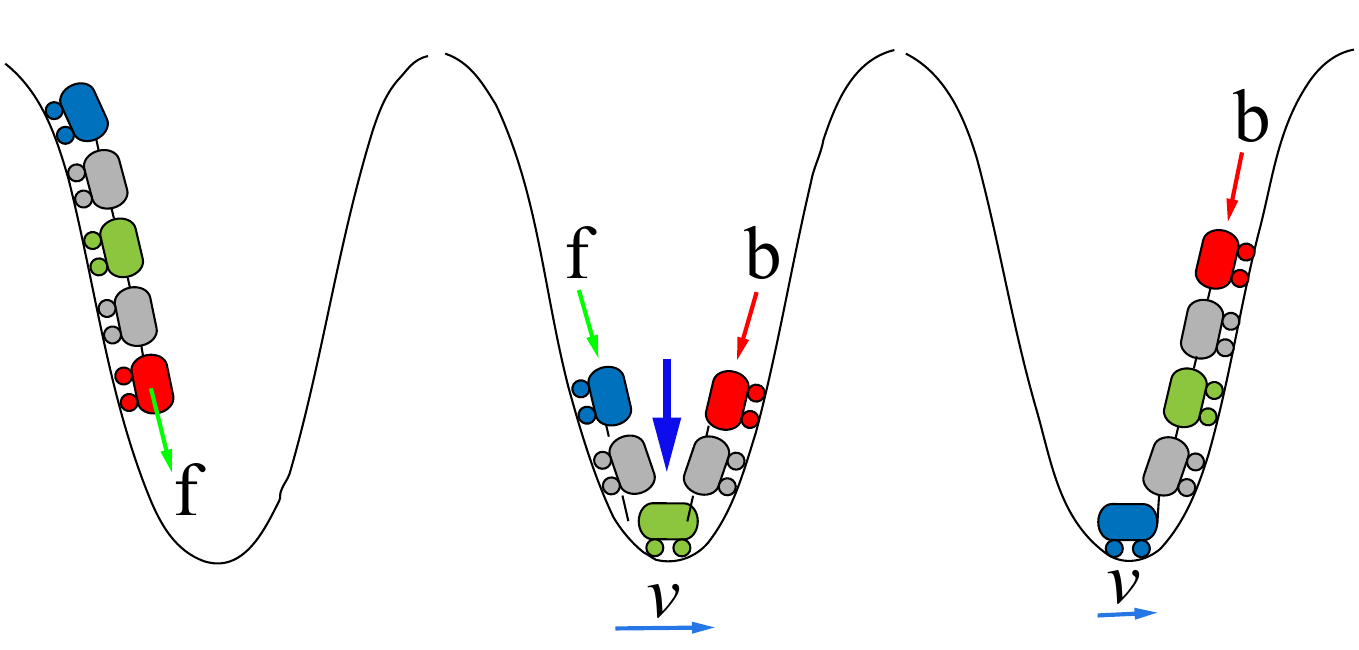}
  \caption{Roller coaster train -- ``valley phase''.  The thick dark blue
    arrow marks the position at which the passengers of the middle wagon feel
    the largest normal force.
    The lengths of the light blue arrows indicate the velocity of the wagon in
    the valley. Arrows at the outer wagons have the same meaning as in
    Fig.~\ref{fig:train_hill}.}
  \label{fig:train_valley}
\end{figure}

At descent the effect is different, see Fig.~\ref{fig:train_valley}.
When the middle wagon is in the ``valley'', the tangential acceleration is zero.
At later times the tangential acceleration will be negative. Thus, the middle
wagon passes the minimum of the track with the highest speed. Passengers in it
feel the strongest normal force. On typical roller coaster tracks accelerations
felt by passengers can be in the range between $2\,g$ and $5\,g$
\cite{Pendrill2020}.

In Sec.~S-II of the supplementary material we show how quantitative
calculations for this model are possible. In the following section we
develop a more sophisticated model consisting of several wagons connected via
springs.

\section{Train  model 2: Wagons connected via springs}
\label{sec:wagons_springs}

In this section we assume that the train consists of several
wagons connected via springs. This model allows the wagons to move relative to
each other. Such motions will (on small scales) always be present since there
will always be some elasticity in the system, though this is unlikely to be
significant in a realistic roller coaster. However, such discrete models are
often used as a pedagogical step towards a continuum model. The model will
be refined in Sec.~\ref{sec:elastic}.

Our model consists of a chain of $N$ identical wagons ($i = 1,2,\ldots,N$) with
mass $m$. Between the masses we assume a coupling with a (very) hard spring of
spring constant $k$ and relaxed length $s_0$.   We consider the spring to be
``hard'' if the distance $d$ between two consecutive wagons is not changed
more than 10\% by the weight of one wagon, i.e., $k > \SI{50}{\kilo\newton
  \per  \metre}$ for a wagon of $m=\SI{500}{\kilo\gram}$. The
situation is very similar to that
depicted in Figs.~\ref{fig:train_hill} and \ref{fig:train_valley}.

In addition to the gravitational interaction the potential energy
must contain the harmonic contribution of the springs, which leads to
\begin{multline}
  U\left ( \{x_i\},\{z_i\}\right ) = m\,g\,\sum_{i=1}^{N}z_i \\
  + \frac{1}{2}\,k\sum_{i=1}^{N-1}\left ( \sqrt{\left ( x_{i+1}-x_i\right )^2
      + \left ( z_{i+1}-z_i\right )^2 }- s_0\right )^2
  \;. \label{eq:discrete_pot}
\end{multline}
The curly braces $\{\cdot\}$ indicate that all $x_i$ and $z_i$ appear in
the potential. Similarly, the total kinetic energy is a sum of the
kinetic energies of all wagons,
\begin{align}
  T\left ( \{\dot{x}_i\},\{\dot{z}_i\}\right ) = \frac{1}{2}\,m\sum_{i=1}^{N}
  \left ( \dot{x}_i{}^2 + \dot{z}_i{}^2\right ) \; ,
  \label{eq:discrete_kin}
\end{align}
and the constraints can be obtained by generalizing those given already
in Eq.~\eqref{eq:constraint_point_particle} for a single point particle, where
we now have $N$ conditions, one for each wagon.
\begin{equation}
  f_i = z_i - h(x_i) = 0
  \label{eq:constraints_Nwagons}
\end{equation}
The derivation of the equations of motion and the normal
forces follows the same steps as already done for a single point mass
in Sec.~\ref{sec:point_particle}. The calculations are somewhat lengthy, and
are provided in Sec.~S-I~A of the supplementary material; here we solve the
equations numerically. We expect that the continuous limit is achieved for
large $N$ and small $s_0$ with
\begin{equation*}
  M = \sum_{i=1}^{N}m = N\,m \quad \text{and} \quad L=N\,s_0 \;.
\end{equation*}

\begin{figure}[tbp]
  \centering
  \includegraphics[width=\columnwidth]{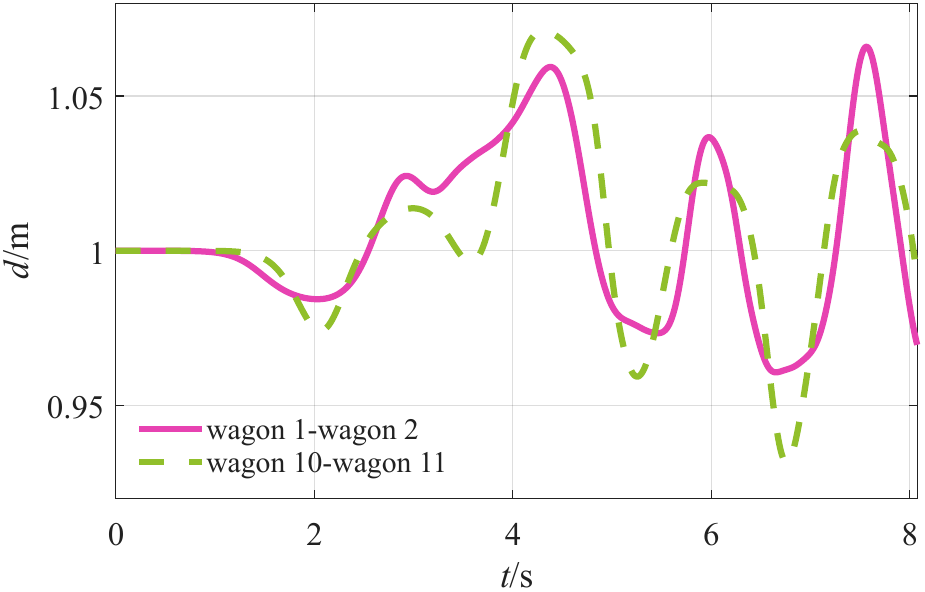}
  \caption{Distance $d$ between the wagons 1 and 2 as well as 10 and
    11 as a function of time. Oscillations initiated by
    the springs are clearly visible. Each wagons has a mass of
    $m = \SI{500}{\kilo\gram}$. Their equilibrium distance is
    $s_0 = \SI{1}{\metre}$, and the spring constant is
    $k=\SI{100}{\kilo\newton \per \metre}$. The track has the Gaussian shape
    \eqref{eq:gaussian} with a height of $H = \SI{20}{\metre}$ and a width
    parameter of $a = \sqrt{\ln 2}/\SI{10}{\metre}$.}
  \label{fig:springs_distances}
\end{figure}

Fig.~\ref{fig:springs_distances} shows the distance between two
pairs of wagons for a train of 11 wagons, each having a mass $500$kg, and with
$k = 100 \,{\rm kN\,m^{-1}}$ and $s_0 = 1$m, which satisfies the
``hard spring'' condition.
As can clearly be seen, the maximum dilation of two springs is $\SI{10}
{\centi\metre}$, while complicated relative motions of the wagons are
possible. We can observe oscillations initiated by the springs, which are
still visible at the end of the track, i.e., some energy is transferred into
purely relative (``vibrational'')  motion of the wagons.

\begin{figure}[tbp]
  \centering
  \includegraphics[width=\columnwidth]{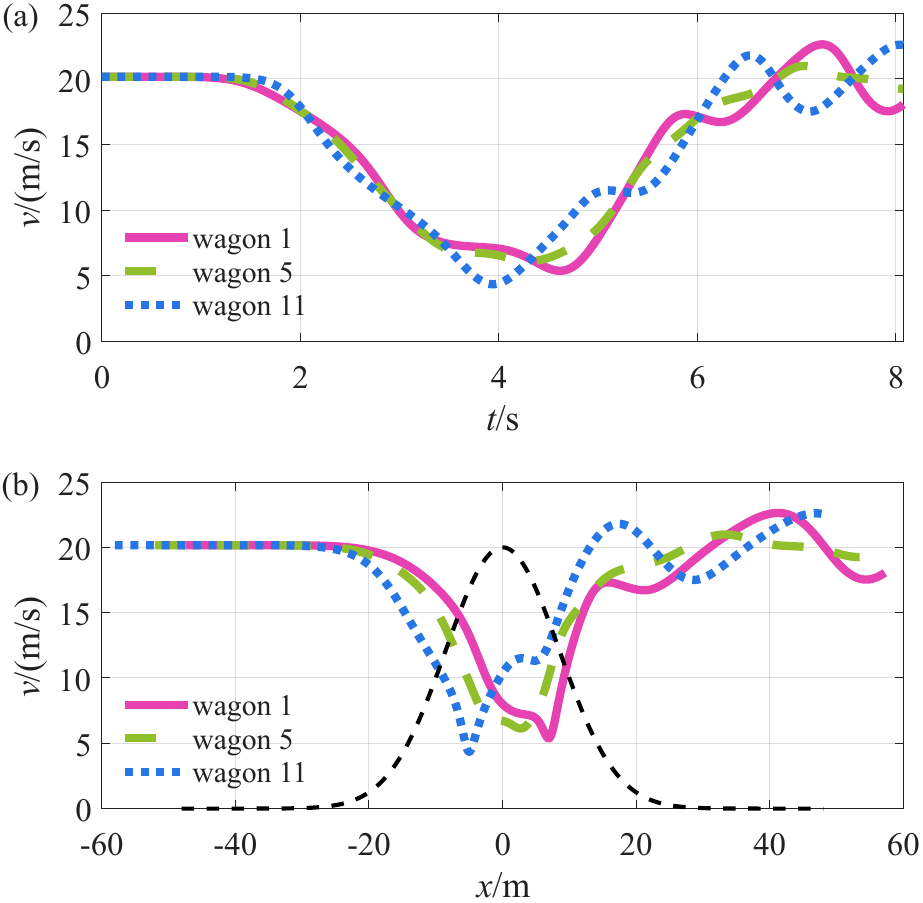}
  \caption{Velocities of front, middle, and rear wagons on a Gaussian track.
    The velocities are shown as (a) a function of time and (b) of the position
    on the track. In (b) the Gaussian shape of the track is shown by the
    dashed line. The parameters are the same as in Fig.\
    \ref{fig:springs_distances}. See text for an explanation of the motions.}
  \label{fig:springs_velocities}
\end{figure}

The velocities of the first, middle, and last wagon are shown as a function
of time in Fig.~\ref{fig:springs_velocities}a.
It might be surprising that the last wagon assumes its lowest velocity before
the first one since it passes the maximum at a later time. However, this can
be understood if we look at Fig.~\ref{fig:springs_velocities}b, in which the
velocities are shown as a function of the track position. The last
wagon is decelerated by the gravitational force on its way up
the hill. However, this does not necessarily lead to a deceleration of
the other wagons, which have already passed the hill. The spring in front
of the last wagon is extended, and the other wagons can still be accelerated
on their way down the hill. After some time the rear-most spring is extended
enough to accelerate the last wagon. As a consequence it reaches its lowest
velocity before it passes the hilltop.

By contrast, the first wagon reaches its minimum velocity only after the
hilltop. Due to the extending springs,it takes some time for it to be affected
by the other ascending wagons. Once the springs are more
extended the first wagon decelerates even though it is on its way downward.
The middle wagon`s velocity reaches two minima owing to the fact that it is
coupled to springs on both sides. 

\begin{figure}[tbp]
  \centering
  \includegraphics[width=\columnwidth]{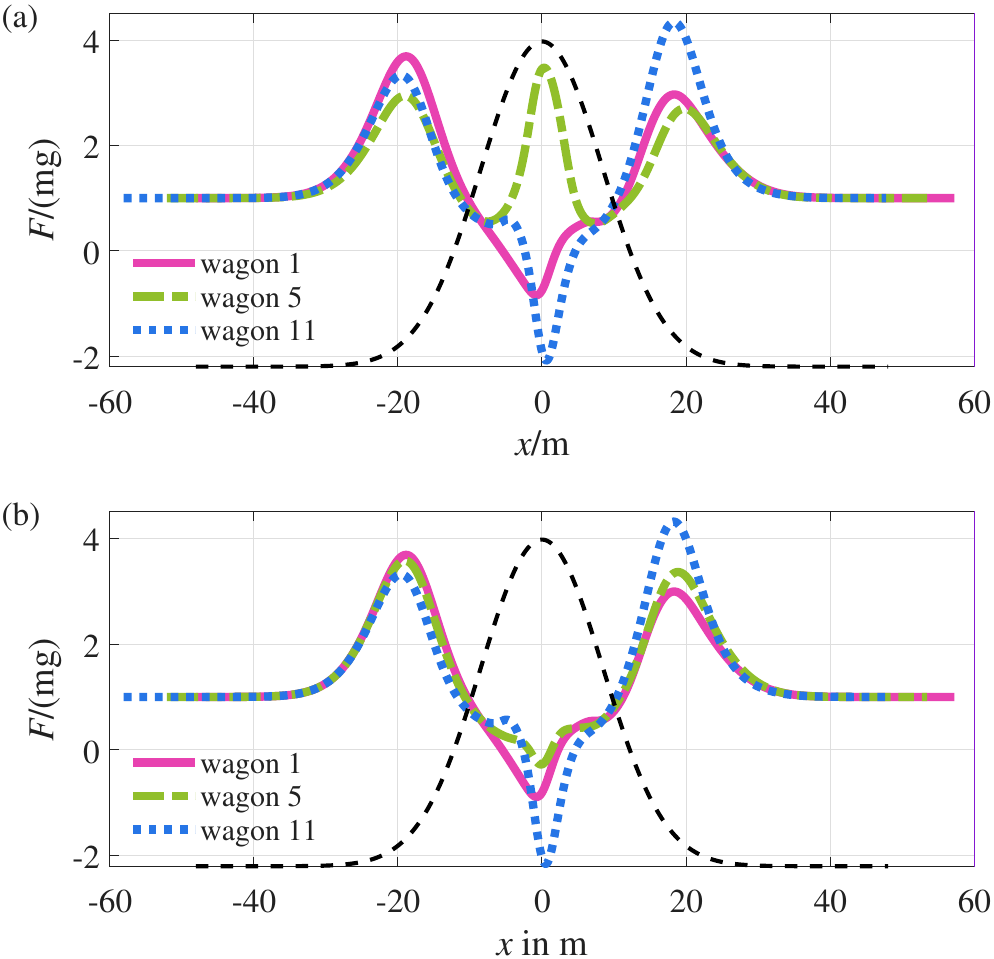}
  \caption{(a) Normal forces acting on selected wagons for a train of
    $N=11$ wagons on a Gaussian track and (b) normal forces acting on the
    passengers in the same wagons. The shape of the track is indicated by
    the dashed line. The coordinate system is chosen such that a force
    acting perpendicularly away from the track is denoted as positive,
    while one acting perpendicularly toward the track is negative.  We
    note that the middle wagon needs an upwards normal force at the
    hilltop to stay on the track, but the passengers in that wagon still
    feel an airtime since the springs act on the wagons, but not on the
    passangers.  The parameters are the same as in Fig.\
    \ref{fig:springs_distances}.}
  \label{fig:springs_normal_forces}
\end{figure}

The normal forces on the wagons by the track are shown in
Fig.~\ref{fig:springs_normal_forces}a.
The outer wagons (1 and 11) need to be pulled onto the track to remain in
contact with the track near the hilltop. The gravitational force would not be
enough to do this. However, this is completely different for the middle wagon
(number 5), which is held onto the track by the surrounding spring forces.

Let us assume the wagons in this model contain passengers.  The springs
influence the motion of the cars, but do not directly act on the passengers.
As such, the passengers may experience airtime even if the wagon itself does
not need to be held in contact with the track. Indeed, this phase of airtime
is revealed by removing the forces exerted by the springs in the calculation
of the normal force since they have no direct effect on the passenger. Then we
obtain the results shown in Fig.~\ref{fig:springs_normal_forces}b. All
passengers feel an airtime. Passengers in the middle wagon experience the
weakest airtime (the restraining bar's force which holds them in their seat
is comparatively small), whereas the passengers in the rear wagon experience
the most intense airtime.  This is consistent with expectations from
Sec.~\ref{sec:fixed_length}.  A detailed quantitative example, which requires
elaborate calculations is given in Sec.~S-III~B of the supplementary material.  

\section{Train model 3: Continuous train of variable length with defined
  elasticity}
\label{sec:elastic}

\subsection{Equations of motion}
\label{sec:elastic_eqs_motion}

The discrete train model we considered in the previous section can be further
extended to a continuous model, which assumes a homogeneous (but still hard
elasticity of the whole train. Instead of wagons connected via
springs, we introduce a continuum limit, in which the number of wagons
approaches infinity while their length vanishes. This models a very long train
of very short wagons. Of course, this is not intended to describe a realistic
roller coaster train (which would have a vanishingly small elasticity).
However, the comparison to the previous model allows for a nice demonstration
of how the different formalisms are applied.

The introduction of the continuous elastic model can be achieved if we
start with the potential \eqref{eq:discrete_pot} and the kinetic energy
\eqref{eq:discrete_kin} of the previous model, substitute $z_i=h(x_i)$ , and
introduce the continuum limit via
{ \allowdisplaybreaks
  \begin{subequations}
    \begin{align}
      x_i &\to x(l,t) \; , \label{eq:continuum_a} \\
      s_0 &= \mathrm{d} l \; , \\
      x_{i+1} - x_i &\to \frac{\partial x}{\partial l} \, s_0 =
                      \frac{\partial x}{\partial l} \, \mathrm{d} l \; , \\
      h(x_{i+1}) - h(x_i)
          &\to \frac{\partial h}{\partial l} \, s_0
            = h'(x) \frac{\partial x}{\partial l} \,\mathrm{d} l \; , \\
      m &= \varrho_L \,s_0 = \varrho_L \,\mathrm{d} l \; , \\
      k &= \frac{\Phi}{s_0} \label{eq:continuum_f} \; ,
    \end{align}
  \end{subequations}
}%
where we describe each (matter) point on the train by a variable $l$ with
the property $0\leq l \leq L$. In this definition, $L$ is the length of a
relaxed train on a horizontal track. The actual position of a single
(matter) point of the train at an arbitrary time $t$ is given by
$x(l,t)$.\footnote{The matter points of the train are described in
  terms of the variable $l$, which is independent of the actual position
  of the train, and is always defined for the relaxed train on a horizontal
  track.  Thus, $l$ can only assume values between $0$ and $L$. The actual
  length of the train on the track can assume values larger
  or smaller than $L$.}
In the continuum limit, the distance between two wagons in the relaxed
state becomes $s_0 = \mathrm{d} l$. The mass has to be replaced by a mass
density (per length) $\varrho_L$. The spring constant multiplied by the unit
length is denoted $\Phi$. Then the limit is achieved by $s_0 \to 0$ and
$N \to \infty$ with a constant total mass $M = \varrho_\text{L}\,L$ and
constant $\Phi$.

Details of the derivation are given in Sec.~S-I~B of the supplementary material.
The most important result is the Lagrangian function
\begin{subequations}
  \begin{equation}
    \mathcal{L} \left ( x , \frac{\partial x}{\partial t} , \frac{\partial x}
      {\partial l} \right )
    = \int_{0}^{L} L \left ( x , \frac{\partial x}{\partial t} ,
      \frac{\partial x}{\partial l} \right ) \, \mathrm{d} l
    \label{eq:elastic_Lfunction}
  \end{equation}
  with a Lagrangian density  
  \begin{multline}
    L \left ( x , \frac{\partial x}{\partial t} , \frac{\partial x}{\partial l}
    \right )
    = \frac{1}{2}\,\varrho_L \left (  1 + h'(x){}^2 \right ) \left (
      \frac{\partial x}{\partial t} \right )^2 \\
    - \varrho_L\,g\, h(x) - \frac{1}{2}\,\Phi\, \left ( \sqrt{1 + h'(x)^2}
    \, \frac{\partial x}{\partial l} \, - 1\right )^2 \; ,
    \label{eq:elastic_Ldensity}
  \end{multline}
\end{subequations}
where $x=x(l,t)$ is a function of $l$ and $t$.

The Lagrangian equation of motion of the second kind can be obtained with
the calculus of variations as is done in detail in Sec.~S-I~B of the
supplementary material. It reads
\begin{multline}
  \frac{\partial^2 x}{\partial t^2} = -\frac{h'(x)h''(x)}{1 + h'(x){}^2}
  \left ( \frac{\partial x}{\partial t} \right )^2
  - \frac{g h'(x)}{1 + h'(x){}^2} \\
  +\frac{\Phi}{\varrho_L}\, \frac{h'(x)h''(x)}{1 + h'(x){}^2} \left (
  \frac{\partial x}{\partial l} \right )^2
  + \frac{\Phi}{\varrho_L}\, \frac{\partial^2 x}{\partial l^2} \; .
  \label{eq:elastic_eqs_motion}
\end{multline}
Note in particular that for a train with fixed length, i.e., for which
$ \partial x / \partial l = \partial^2 x / \partial l^2 = 0$, the
equation of motion \eqref{eq:elastic_eqs_motion} reduces to the one
already obtained for the point particle
Eq.~\eqref{eq:eqs_motion_pointx}.

\subsection{Examples for the train on a Gaussian track}

\begin{figure}[tbp]
  \centering
  \includegraphics[width=\columnwidth]{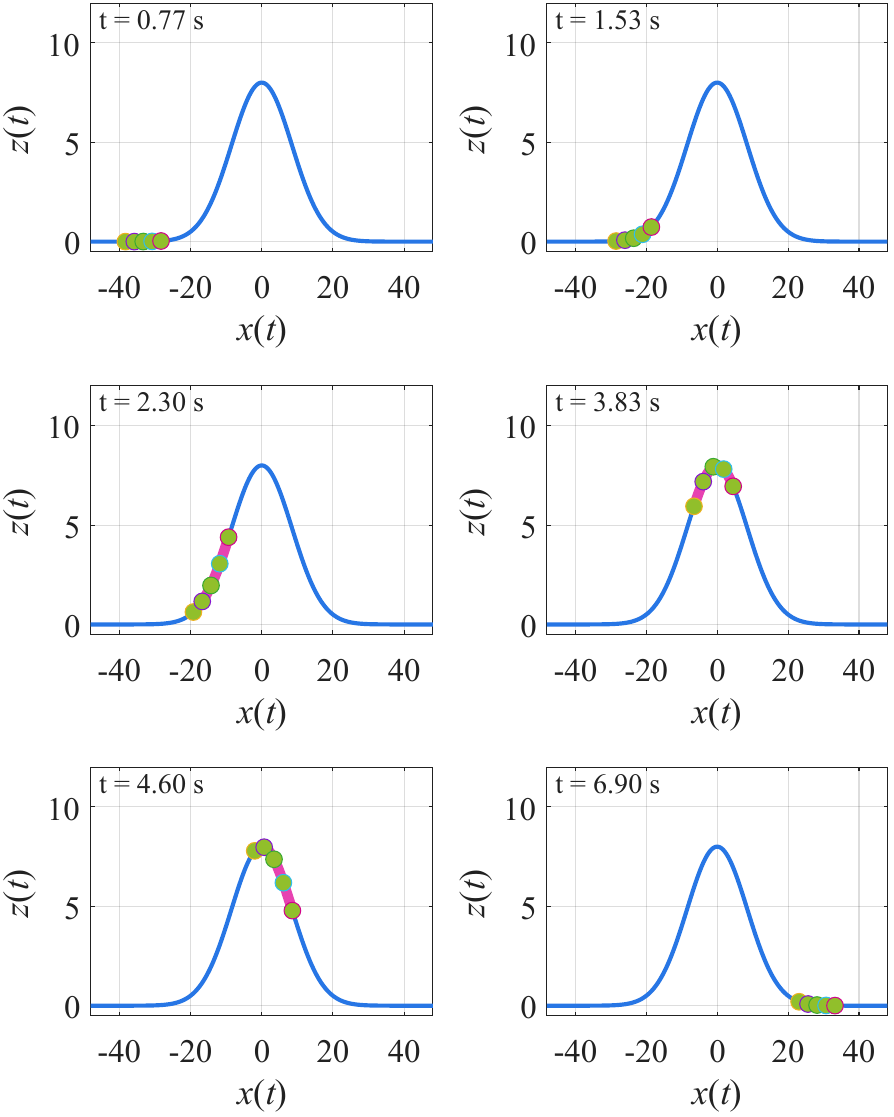}
  \caption{Elastic roller coaster train on a Gaussian track at
    different times. The green filled circles correspond to matter points
    on the train, which are equidistance when the train is in its relaxed
    state. The track is a Gaussian of the form given in
    Eq.~\eqref{eq:gaussian} with a height of $H = \SI{8}{\metre}$ and a
    half width at half maximum of $\SI{10}{\metre}$, i.e., $a = \sqrt{\ln
      2}/\SI{10}{\metre}$. The initial velocity is $\dot{x}_0 = 1.018 \times
    \dot{x}_{0,\mathrm{min}}$ with $\dot{x}_{0,\mathrm{min}}$ from
    Eq.~\eqref{eq:dotxmin}.}
  \label{fig:elastic_distances}
\end{figure}

A numerical solution of the equation of motion \eqref{eq:elastic_eqs_motion}
can be obtained quite easily with the iteration equation
derived in Sec.~S-I~C of the supplementary material. An example for
the same Gaussian track we already discussed in Secs.~\ref{sec:point_particle}
and \ref{sec:wagons_springs} is given in
Fig.~\ref{fig:elastic_distances},
in which the train starts (at time $t=0$) far away from the hill.

When the train was modeled as wagons connected by springs, we learned in
Sec.~\ref{sec:wagons_springs} that the distance between the first two wagons
first shrinks when the train starts to ascend on the Gaussian track and later
increases when more wagons are on the steep slope. With increasing time
this results in an oscillation mediated by the springs. In
Fig.~\ref{fig:elastic_distances} we observe a similar effect. The green points
mark points on the train that are a fixed distance apart in the relaxed
state, i.e., at the start of the motion when the train is on a straight
horizontal line track. However, at $t=\SI{1.53}{\second}$ we observe that the
distance between the first two points increases. This is amplified at later
times as can be seen for $t=\SI{2.30}{\second}$.
Around the hilltop the points get closer to each other. When the train is
beyond the hilltop the distances increase again, which is identical to the
behavior in the discrete train model in Sec.~\ref{sec:wagons_springs}. The
train compresses again once the front segments have reached the flat part of
the track, but the rear segments are still accelerating down the hill.

When all segments have reached the flat part some residual oscillations remain
in the train. This means the train does not return to its original
fixed length. As in the discrete model of Sec.~\ref{sec:wagons_springs}
some of the energy is transferred into an internal vibrational
motion.\footnote{In a realistic train the relative motion would be damped,
  which means that after long time the train would return to its initial
  relaxed form but with less speed than at the beginning of the track. To
  account for this effect, the model can be easily extended by including a
  dissipation function into the Lagrangian.}

\subsection{Normal forces acting on the passengers}

Since we had to use the Lagrangian equations of the second kind to derive
the equation of motion for the elastic continuous model, the extraction
of the normal forces is a bit more complicated. This is done in Sec.~S-I~D
of the supplementary material. We obtain the normal force components
experienced by a passenger in the train 
\begin{subequations}
  \begin{align}
    F_{N,x} &= -m\frac{h'' h'}{1+h'^2}\, \dot{x}^2 \; , \\
    F_{N,z} &= m\frac{h''}{1+h'^2}\, \dot{x}^2 +mg \; .
  \end{align}
\end{subequations}

\begin{figure}[tbp]
  \centering
  \includegraphics[width=\columnwidth]{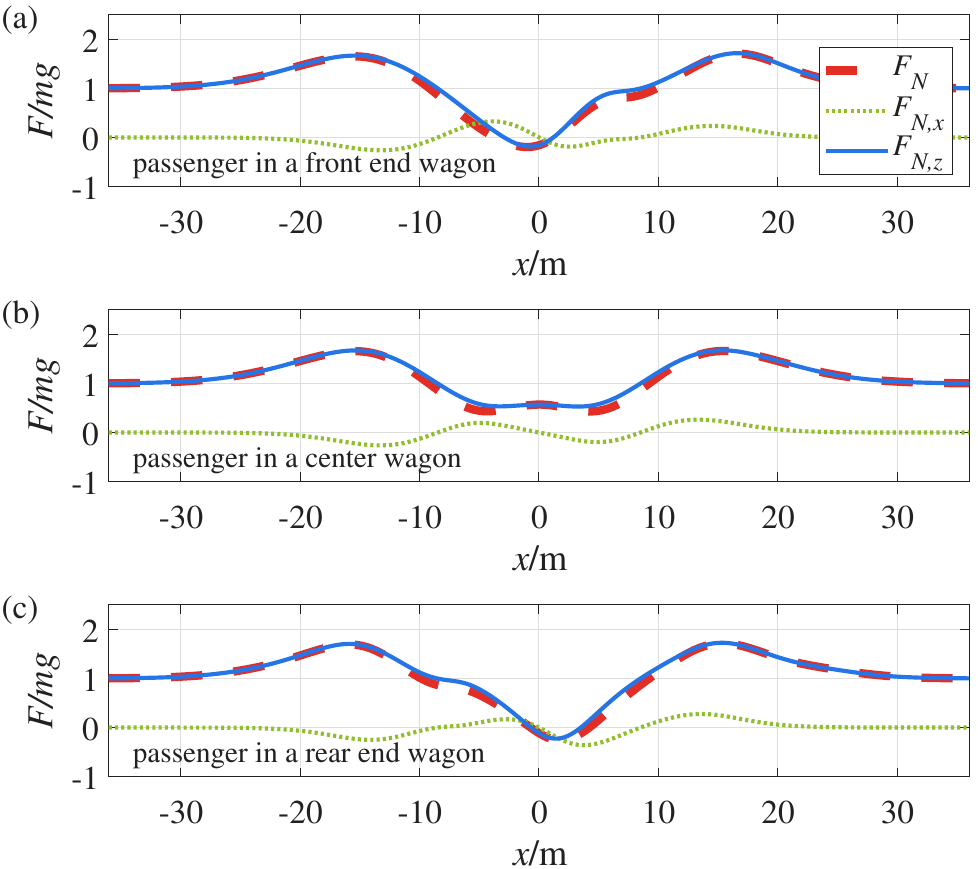}
  \caption{Components ($F_{N,x},\,F_{N,z}$) of the normal force acting on a
    passenger of a roller coaster train on a Gaussian track as determined in
    the continuous elastic model, (a) passenger in the front part, (b)
    passenger in the center part, and (c) passenger in the rear part. The
    parameters are the same as in Fig.~\ref{fig:elastic_distances}.}
  \label{fig:elastic_normal_accelerations}
\end{figure}

For the Gaussian track, numerically calculated values are given in
Fig.~\ref{fig:elastic_normal_accelerations}.
In comparison with the discrete model of Sec.~\ref{sec:wagons_springs} we
observe that the accelerations are less symmetric around the hilltop. Due
to the uniform elasticity, energy is transferred more quickly
into the oscillations of the train's length as compared to a train with
only a few hard springs. The stronger oscillations result in a more complex
motion.

However, the qualitative progression of the normal forces already observed in
Sec.~\ref{sec:wagons_springs} is confirmed once again. Around the
hilltop the security bar of the wagon is required to keep the passengers
in their seatsand and the passengers feel an airtime.
As in the discrete model a passenger in the middle wagon feels a weaker
force. 

\section{Conclusion}
\label{sec:conclusion}

The roller coaster and elastic extensions of it are an ideal context to
discuss different approaches of classical mechanics. It is particularly
instructive to have students investigate this case since nontrivial physical
effects that can be experienced with their own body or measured by a smart
phone sensor can be well understood. Newtonian mechanics and Lagrangian
mechanics of the first kind for a point particle offer simple access to the
forces felt by passengers. However, a full understanding is only possible if
the finite length of the train is considered. The influence of the position
in the train can already qualitatively be discussed with a fixed-length train.
Additional effects appear if the elasticity of the train is added to the
model. This can be done by a train consisting of a chain of point masses
connected via springs or in a continuous model. For the latter we used
Lagrangian mechanics of the second kind with a Lagrange density of an elastic
band.

On the conceptual side, the roller coaster can be used to
show students how making a model sophisticated adds more and more physical
effects and insights that are not visible in the simpler models. 

Last but not least, we learn from the roller coaster that constraints can lead
to effects which are intuitive for learners.  The fact that the train is
constrained to stay on the track leads to experiences like ``airtime'' which
are exciting and attractive to thrill-seekers.

\section*{Acknowledgements}

The authors would like to thank a reviewer as well as the editors for
valuable comments and suggestions to improve the article.

\section*{Author Declarations}

\subsection*{Conflict of Interest}

The authors have no conflicts to disclose.

\subsection*{Author Contributions}

All authors contributed equally to this work.

\clearpage

\onecolumngrid

{\LARGE \textbf{Supplementary material}}

\renewcommand{\theequation}{S\arabic{equation}}
\setcounter{equation}{0}

\renewcommand{\thefigure}{S\arabic{figure}}
\setcounter{figure}{0}

\renewcommand{\thesection}{S-\Roman{section}}
\setcounter{section}{0}

\section{Additional calculations steps in the derivations}

\subsection{Derivation of the equations of motion for the train model with
  wagons connected via springs}
\label{sec:app_springs}

To derive the equations of motion required in Sec.~IV of the main text we
first calculate the derivatives of the constraints~(13)
\begin{subequations}
  \begin{align}
    \dot{f}_i &= \dot{z}_i - \frac{\partial h}{\partial x_i}\,\dot{x}_i = 0
                = \dot{z}_i
  - h'\,\dot{x}_i \;, \\
  \ddot{f}_i &= \ddot{z}_i - \frac{\partial^2 h}{\partial x_i^2}\,\dot{x}_i{}^2
               - \frac{\partial h}{\partial x_i}\,\ddot{x}_i = 0 = \ddot{z}_i
               - h''\,\dot{x}_i{}^2
               - h'\,\ddot{x}_i \; . \label{eq:constraints_ddzi}
  \end{align}
\end{subequations}
The $2N$ equations of motion (Lagrange equations of the first kind) read
\begin{subequations}
  \begin{align}
    \frac{\mathrm{d}}{\mathrm{d} t}\,\frac{\partial \mathcal{L}}{\partial \dot{x}_i}
    - \frac{\partial \mathcal{L}}{\partial x_i}
    &= m\,\ddot{x}_i + \frac{\partial U}{\partial x_i}
      = \sum_{n=1}^{N} \lambda_n\,\frac{\partial f_n}{\partial x_i} 
    = \sum_{n=1}^{N} \delta_{ni}\,\lambda_n\,\frac{\partial f_n}{\partial x_i}
    = \lambda_i\,\frac{\partial f_i}{\partial x_i} \;, \\
    \frac{\mathrm{d}}{\mathrm{d} t}\,\frac{\partial \mathcal{L}}
    {\partial \dot{z}_i}
    - \frac{\partial \mathcal{L}}{\partial z_i}
    &= m\,\ddot{z}_i + \frac{\partial U}{\partial z_i}
      = \sum_{n=1}^{N} \lambda_n\,\frac{\partial f_n}{\partial z_i} 
    = \sum_{n=1}^{N} \delta_{ni}\,\lambda_n\,\frac{\partial f_n}{\partial z_i}
      = \lambda_i\,\frac{\partial f_i}{\partial z_i} \; ,
  \end{align}
\end{subequations}
where the Kronecker delta $\delta_{ni}$ was used since each $f_i$ only depends
on $x_i$ and not on the other variables $x_n, n \neq i$.
The equations of motion are not independent since they are coupled due
to the terms $\partial U/\partial x_i$ and $\partial U/\partial z_i$. This is
a consequence of the next-neighbor coupling of the wagons with the springs.
Since the first and last wagon only have one next neighbor, we have to treat
them separately. Using the potential (11) and the kinetic energy (12) we
obtain for the first wagon ($i=1$)
\begin{subequations}
  \begin{align}
    m\,\ddot{x}_1 &= -k\,\frac{\sqrt{\left ( x_2-x_1\right )^2 + \left ( z_2-z_1\right )^2} -s_0}
                    {\sqrt{\left ( x_2-x_1\right )^2 + \left ( z_2-z_1\right )^2}}\,\left ( x_2-x_1\right )
                    - \lambda_1\,h'\left ( x_1\right ) \;,
                    \label{eq:springs_eqsm_x1} \\
    m\,\ddot{z}_1 &= -k\,\frac{\sqrt{\left ( x_2-x_1\right )^2 + \left ( z_2-z_1\right )^2} -s_0}
                    {\sqrt{\left ( x_2-x_1\right )^2 + \left ( z_2-z_1\right )^2}} \left ( z_2
                    - z_1\right ) -m\,g + \lambda_1 \; , \label{eq:springs_eqsm_z1}
  \end{align}
\end{subequations}
and for the last ($i=N$)
\begin{subequations}
  \begin{align}
    m\,\ddot{x}_N &= -k \frac{\sqrt{\left ( x_N - x_{N-1}\right )^2
                    + \left ( z_N - z_{N-1}\right )^2} - s_0}{\sqrt{\left ( x_N - x_{N-1}\right )^2
                    + \left ( z_N - z_{N-1}\right )^2}}\, \left ( x_N - x_{N-1}\right ) - \lambda_N\,h'\left ( x_N\right ) \;,\\
		m\,\ddot{z}_N &= -k \frac{\sqrt{\left ( x_N - x_{N-1}\right )^2
                    + \left ( z_N - z_{N-1}\right )^2} - s_0}{\sqrt{\left ( x_N - x_{N-1}\right )^2
                    + \left ( z_N - z_{N-1}\right )^2}}\, \left ( z_N - z_{N-1}\right ) - m\,g + \lambda_N\,.
  \end{align}
\end{subequations}
For all other wagons ($i=2,3,\ldots,N-1$) we obtain
\begin{subequations}
    \begin{align}
      m\,\ddot{x}_i
      &= -k\, \frac{\sqrt{\left ( x_i - x_{i-1}\right )^2 + \left ( z_i- z_{i-1}\right )^2} - s_0}
        {\sqrt{\left ( x_i - x_{i-1}\right )^2 + \left ( z_i - z_{i-1}\right )^2}}
        \,\left ( x_i -x_{i-1}\right ) 
      + k\, \frac{\sqrt{\left ( x_{i+1} - x_i \right )^2 + \left ( z_{i+1}- z_i \right )^2} - s_0}
        {\sqrt{\left ( x_{i+1} - x_i \right )^2 + \left ( z_{i+1} - z_i \right )^2}} \,\left ( x_{i+1}
        -x_i\right ) \notag \\
      &\quad+ \lambda_i\,h'\left ( x_i\right ) \;, \\
      m\,\ddot{z}_i
      &= -k\, \frac{\sqrt{\left ( x_i - x_{i-1}\right )^2 + \left ( z_i- z_{i-1}\right )^2} - s_0}
        {\sqrt{\left ( x_i - x_{i-1}\right )^2 + \left ( z_i - z_{i-1}\right )^2}}\,\left ( z_i
        - z_{i-1} \right ) 
      +k\, \frac{\sqrt{\left ( x_{i+1} - x_i \right )^2 + \left ( z_{i+1}- z_i \right )^2} - s_0}
        {\sqrt{\left ( x_{i+1}- x_i \right )^2 + \left ( z_{i+1} - z_i \right )^2}}\,
        \left ( z_{i+1} - z_i\right ) \notag \\
      &\quad -m\,g + \lambda_i\,.
    \end{align}
\end{subequations}
In the following we abbreviate the terms containing square roots with
\begin{align*}
  \Delta_{i,j} &= \sqrt{\left ( x_i-x_j\right )^2 + \left ( z_i-z_j\right )^2} \; ,~~
  S_{i,j} = \frac{\Delta_{i,j}-s_0}{\Delta_{i,j}} \; .
\end{align*}
The Lagrangian multipliers $\lambda_i$ can be determined with the equations
of motion and the constraints. To do so, we combine the equations of motion
for $x_i$ and $z_i$ via the corresponding constraint
\eqref{eq:constraints_ddzi} and solve the resulting equation for $\lambda_i$.
We obtain
\begin{subequations}
    \begin{align}
      \lambda_1 &= \frac{m\,g + m\,h''\,\dot{x}_1{}^2 + k\,S_{2,1}
                  \left [ h'\left ( x_2-x_1\right ) - \left ( z_2-z_1\right )\right ]}{1+h'{}^2} \;, \\
      \lambda_N &= \frac{m\,g + m\,h''\,\dot{x}_N{}^2 - k\,S_{N,N-1}\left [ h'
                  \left ( x_N-x_{N-1}\right ) - \left ( z_N - z_{N-1}\right )\right ]}{ 1+h'{}^2} \;, \\
      &\text{and for all $2 \le i \le N-1$} \notag \\
      \lambda_i &= \frac{m\,g + m\,h''\,\dot{x}_i{}^2 - k\,S_{i,i-1}\,
                  \left [ h'\left ( x_i-x_{i-1}\right ) - \left ( z_i - z_{i-1}\right )\right ] 
                  \,+k\,S_{i+1,i} \,\left [ h'\left ( x_{i+1}-x_i\right ) - \left ( z_{i+1}
                      -z_i\right )\right ]}{1+h'{}^2} \; .
    \end{align}
\end{subequations}

\subsection{Derivation of the equations of motion in the model of a
  continuous train with defined elasticity}
\label{sec:app_elastic_continous}
With the relation $z_i=h(x_i)$ we introduce the independent variables $x_i$,
which leads according to Eqs.~(12) and (11) to the expressions for the
potential and kinetic energies,
\begin{align}
  U\left ( \{x_i\}\right )
  &= m\,g\,\sum_{i=1}^{N} h(x_i) + \frac{1}{2}\,k\sum_{i=1}^{N-1}\left ( \sqrt{\left ( x_{i+1}-x_i\right )^2
    + \left (  h(x_{i+1})-h(x_i)\right )^2 }- s_0\right )^2 \; , \\
  T\left ( \{\dot{x}_i\},\{\dot{z}_i\}\right )
  &= \frac{1}{2}\,m\sum_{i=1}^{N}\left ( \dot{x}_i{}^2 + \dot{h}(x_i){}^2 \right ) = \frac{1}{2}\,m\sum_{i=1}^{N}\left (  1 + h'(x_i){}^2 \right )
    \dot{x}_i{}^2 \; .
    \label{eq:kinetic_continuous}
\end{align}

With these definitions we write the potential as
\allowdisplaybreaks
\begin{align}
      U\left ( \{x_i\}\right )
      &= m\,g\,\sum_{i=1}^{N} h(x_i)+\frac{1}{2}\,k\sum_{i=1}^{N-1} \left (
        \sqrt{\left ( x_{i+1}-x_i\right )^2 + \left ( h(x_{i+1})-h(x_i) \right )^2 }- s_0\right )^2
        \notag \\
      &=  \varrho_L\,g\,\sum_{i=1}^{N} h(x_i) s_0 + \frac{1}{2}\,k\sum_{i=1}^{N-1} \left (
        \sqrt{\left ( \frac{\partial x}{\partial l} s_0 \right )^2 + \left ( h'(x)
        \frac{\partial x}{\partial l} \,s_0 \right )^2 }- s_0\right )^2 \notag \\
      &=  \varrho_L\,g\,\sum_{i=1}^{N} h(x_i) s_0 + \frac{1}{2}\,k\sum_{i=1}^{N-1}
        \left ( \sqrt{1 + h'(x)^2 }\, \frac{\partial x}{\partial l} \, s_0 - s_0\right )^2
        \notag \\
      &=  \varrho_L\,g\,\sum_{i=1}^{N} h(x_i) s_0 + \frac{1}{2}\,\Phi\,
        \sum_{i=1}^{N-1} \left ( \sqrt{1 + h'(x)^2 }\, \frac{\partial x}{\partial l}
        \, - 1\right )^2 s_0 \;.
\end{align}
This leads for a train of length $L$ in the continuum limit of Eqs. (14a)-(14f)
with $s_0 = \mathrm{d} l \to 0$ and $N\to\infty$ to 
\begin{align}
    U\left ( x , \frac{\partial x}{\partial l}\right ) &=
    \int_{0}^{L} \bigg \{ \varrho_L\,g\, h(x) 
    + \frac{1}{2}\,\Phi\, \left ( \sqrt{1 + h'(x)^2}\, \frac{\partial x}
      {\partial l} \, - 1 \right )^2 \bigg \} \, \mathrm{d} l \; .
  \end{align}
Analogously we find for the kinetic energy the term
\begin{align}
  T\left ( x , \frac{\partial x}{\partial t} \right )
  &= \frac{1}{2}\,m\sum_{i=1}^{N}\left (  1 + h'(x_i){}^2 \right ) \dot{x}_i{}^2
    = \frac{1}{2}\,\varrho_L \sum_{i=1}^{N}\left (  1 + h'(x_i){}^2 \right )
    \dot{x}_i{}^2 \, s_0 = \int_{0}^{L} \frac{1}{2}\,\varrho_L \left (  1 + h'(x){}^2 \right )
    \left ( \frac{\partial x}{\partial t} \right )^2 \, \mathrm{d} l \; .
\end{align}
These expressions are combined to the Lagrange function
(15a) with the Lagrange density (15b).

The equations of motion are derived with the calculus of variations, for which
we first define the action functional
\begin{subequations}
  \begin{equation}
    S =  \int \mathrm{d} t\, \int_{0}^{L} \mathrm{d} l \, L \biggl
    ( x , \frac{\partial x} {\partial l} , \frac{\partial x}{\partial t}
    \biggr )
  \end{equation}
  and calculate its variation,
    \begin{align}
      \delta S &= \int_0^T \mathrm{d} t\, \int_{0}^{L} \mathrm{d} l \, \left \{ L \left ( x
                 + \delta x ,
                 \frac{\partial x}{\partial l} + \frac{\partial \delta x}
                 {\partial l},
                 \frac{\partial x}{\partial t} + \frac{\partial \delta x}
                 {\partial t}
                 \right )
                 - L \left ( x , \frac{\partial x}{\partial l} ,
                 \frac{\partial x}{\partial t}\right ) \right \} \notag \\
               &\overset{\text{linearization}}{\approx}
                 \int_0^T \mathrm{d} t\, \int_{0}^{L} \mathrm{d} l \, \left \{
                 L \left ( x, \frac{\partial x}{\partial l}, \frac{\partial x}
                 {\partial t} \right )
                 + \frac{\partial L}{\partial x} \, \delta x
                 + \frac{\partial L}{\partial \frac{\partial x}{\partial l}}
                 \frac{\partial \delta x}{\partial l}
                 + \frac{\partial L}{\partial \frac{\partial x}{\partial t}}
                 \frac{\partial \delta x}{\partial t}
                 - L \left ( x , \frac{\partial x}{\partial l} ,
                 \frac{\partial x}{\partial t}\right ) \right \} \notag \\
               &= \int_0^T \mathrm{d} t\, \int_{0}^{L} \mathrm{d} l \, \left \{
                 \frac{\partial{L}}
                 {\partial x}\,\delta x + \frac{\partial L}{\partial
                 \frac{\partial x}{\partial l}} \,  \frac{\partial \delta x}
                 {\partial l} + \frac{\partial L}{\partial
                 \frac{\partial x}{\partial t}} \,  \frac{\partial \delta x}
                 {\partial t} \right \} \notag \\
               &\underset{\text{second term}}{\overset{\text{int. by parts}}{=}}
                 \underbrace{\left [ \frac{\partial L}{\partial\frac{\partial x}
                 {\partial l}}\, \delta x \right ]_0^L}_{=0}
                 + \int_0^T \mathrm{d} t\, \int_{0}^{L} \mathrm{d} l \, \left \{
                 \frac{\partial{L}}{\partial x}\,\delta x
                 - \left ( \frac{\partial}{\partial l} \frac{\partial L}
                 {\partial\frac{\partial x}{\partial l}} \right ) \delta x
                 + \frac{\partial L}{\partial \frac{\partial x}{\partial t}}
                 \,  \frac{\partial \delta x}{\partial t} \right \} \notag \\
               &\underset{\text{third term}}{\overset{\text{int. by parts}}{=}}
                 \underbrace{\left [ \frac{\partial L}{\partial\frac{\partial x}
                 {\partial t}}\, \delta x \right ]_0^T}_{=0}
                 + \int_0^T \mathrm{d} t\, \int_{0}^{L} \mathrm{d} l \, \left \{
                 \frac{\partial{L}}{\partial x}\,\delta x
                 - \left ( \frac{\partial}{\partial l} \frac{\partial L}
                 {\partial\frac{\partial x}{\partial l}} \right ) \delta x
                 - \left ( \frac{\partial}{\partial t} \frac{\partial L}
                 {\partial \frac{\partial x}{\partial t}} \right ) \delta x
                 \right \} \notag \\
               &= \int_0^T \mathrm{d} t\, \int_{0}^{L} \mathrm{d} l \, \left \{
                 \frac{\partial{L}}{\partial x}
                 - \frac{\partial}{\partial l} \frac{\partial L}{\partial
                 \frac{\partial x}{\partial l}}
                 - \frac{\partial}{\partial t} \frac{\partial L}{\partial
                 \frac{\partial x}{\partial t}}
                 \right \}  \, \delta x \; ,
    \end{align}
\end{subequations}
from which we read the necessary condition
\begin{equation}
  \frac{\partial{L}}{\partial x}
  - \frac{\partial}{\partial t} \frac{\partial L}{\partial \frac{\partial x}
    {\partial t}} 
  - \frac{\partial}{\partial l} \frac{\partial L}{\partial \frac{\partial x}
    {\partial l}} =0 \; .
  \label{eq:elastic_langrage_pde}
\end{equation}         

To explicitly calculate the equations of motion for our problem, we need
to determine the derivatives which appear in
Eq.~\eqref{eq:elastic_langrage_pde}. They are
\begin{subequations}
  \allowdisplaybreaks
  \begin{align}
    \frac{\partial{L}}{\partial x}
    &= \varrho_L h'(x)h''(x) \left ( \frac{\partial x}{\partial t} \right )^2
      - \varrho_L g h'(x) 
    -\Phi\, h'(x)h''(x) \left ( \frac{\partial x}{\partial l} \right )^2
      + \Phi\, \frac{h'(x)h''(x)}{\sqrt{1 + h'(x)^2}}\,\frac{\partial x}
      {\partial l} \; , \\
    \frac{\partial L}{\partial \frac{\partial x}{\partial l}}
    &= - \Phi\, \left ( \left ( 1 + h'(x)^2 \right )\, \frac{\partial x}
      {\partial l} \, - \sqrt{1 + h'(x)^2} \right ) \; , \\
    \frac{\partial}{\partial l}\frac{\partial L}{\partial \frac{\partial x}
    {\partial l}}
    &= - 2 \Phi\, h'(x)h''(x) \, \left ( \frac{\partial x}{\partial l}
      \right )^2
      - \Phi\, \left ( 1 + h'(x)^2 \right ) \frac{\partial^2 x}{\partial l^2}
      + \Phi\, \frac{h'(x)h''(x)}{\sqrt{1 + h'(x)^2}} \,
      \frac{\partial x}{\partial l} \; , \\
    \frac{\partial L}{\partial\frac{\partial x}{\partial t}}
    &= \varrho_L \left (  1 + h'(x){}^2 \right ) \frac{\partial x}{\partial t}
    \; , \\
    \frac{\partial}{\partial t} \frac{\partial L}{\partial \frac{\partial x}
    {\partial t}}
    &= 2 \varrho_L \, h'(x)h''(x) \left ( \frac{\partial x}{\partial t}
      \right )^2
      + \varrho_L \left (  1 + h'(x){}^2 \right ) \frac{\partial^2 x}
      {\partial t^2}  \; .
  \end{align}
\end{subequations}
Together with the equation of motion \eqref{eq:elastic_langrage_pde} this leads
to 
\begin{equation}
  -\varrho_L h'(x)h''(x) \left ( \frac{\partial x}{\partial t} \right )^2
  - \varrho_L g h'(x) 
  +\Phi\, h'(x)h''(x) \left ( \frac{\partial x}{\partial l} \right )^2 
  + \Phi\, \left ( 1 + h'(x)^2 \right ) \frac{\partial^2 x}{\partial l^2}
  - \varrho_L \left (  1 + h'(x){}^2 \right ) \frac{\partial^2 x}{\partial t^2}
  = 0 \; ,
\end{equation}
from which (16) can be obtained.

\subsection{Discretization of the equations of motion of the continuous
  model for a numerical solution}
\label{sec:app_elastic_discretization}

A numerical solution of the equations of motion (16) requires a
discretization. This partial differential equation can be transformed to an
iteration in the time variable via finite differences. To do so, we introduce
for the first and second derivatives of $x$ with respect to time 
\begin{align*}
  \frac{\partial x}{\partial t}
  &\to \frac{x(l_i,t_{j}) - x(l_i,t_{j-1})}{\Delta t} \; ,\\
  \frac{\partial^2 x}{\partial t^2}
  &\to \frac{x(l_i,t_{j+1}) + x(l_i,t_{j-1}) -2 x(l_{i},t_{j})}{\Delta t^2}
    \; .
\end{align*}
The same concept is applied for the derivatives with respect to the
matter point coordinate $l$ and we obtain
\begin{align*}
  \frac{\partial x}{\partial l}
  &\to \frac{x(l_i,t_j) - x(l_{i-1},t_j)}{\Delta l} \; ,\\
  \frac{\partial^2 x}{\partial l^2}
  &\to \frac{x(l_{i+1},t_j) + x(l_{i-1},t_j) -2 x(l_{i},t_{j})}{\Delta t^2}
    \; .
\end{align*}
Then the discretized version of the differential equation reads
\begin{multline}
  \frac{x(l_i,t_{j+1}) + x(l_i,t_{j-1}) -2 x(l_{i},t_{j})}{\Delta t^2}
  = -\frac{h'(x)h''(x)}{1+h'(x)^2} \left ( \frac{x(l_i,t_{j}) - x(l_i,t_{j-1})}
  {\Delta t} \right )^2 - g \frac{h'(x)}{1+h'(x)^2} \\
  + \frac{\Phi}{\varrho_L} \frac{h'(x)h''(x)}{1+h'(x)^2}
  \left ( \frac{x(l_i,t_j) - x(l_{i-1},t_j)}{\Delta l} \right )^2 
  + \frac{\Phi}{\varrho_L} \frac{x(l_{i+1},t_j) + x(l_{i-1},t_j)
    -2 x(l_{i},t_{j})}{\Delta t^2} \; , 
\end{multline}
which can be rearranged to an iteration equation in time
\begin{multline}
  x(l_i,t_{j+1}) = 2 x(l_{i},t_{j}) - x(l_i,t_{j-1}) 
  - \frac{h'(x)h''(x)}{1+h'(x)^2} \left ( x(l_i,t_{j}) - x(l_i,t_{j-1})
  \right )^2  - g \frac{h'(x)}{1+h'(x)^2} \Delta t^2 \\
  + \frac{\Phi}{\varrho_L} \frac{h'(x)h''(x)}{1+h'(x)^2} \left ( \frac{\Delta t}
  {\Delta l} \right )^2 \left ( x(l_i,t_j) - x(l_{i-1},t_j) \right )^2 
  + \frac{\Phi}{\varrho_L} \left ( \frac{\Delta t}
  {\Delta l} \right )^2 \left ( x(l_{i+1},t_j) + x(l_{i-1},t_j)
  -2 x(l_{i},t_{j}) \right ) \; .
\end{multline}

\subsection{Derivation of the normal forces in the continuous elastic model }
\label{sec:app_elastic_normal}

Since we had to use the Lagrangian equations of motion of the second kind
in Sec.~V~A, we have to extract the normal forces
after we have solved the equations of motion. This can be done in the
following way.

First we have to note that next to the gravitational force a radial force acts
on the passenger, which has the modulus
\begin{equation}
  a_r = \frac{v^2}{r} \label{eq:radial_acc}
\end{equation}
with the velocity $v$ of the passenger on its path and the radius $r$ of a
circle tangential to the path, i.e., its curvature corresponds to that of the
path.

\begin{figure}[tbp]
  \centering
  \includegraphics[width=0.5\columnwidth]{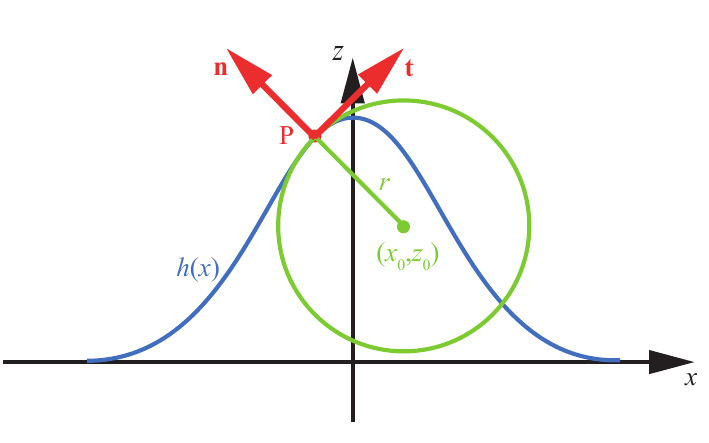}
  \caption{Circle required to determine the radial acceleration. The circle
    touches the path of the passenger and describes the curvature of the path
    at point P correctly.}
  \label{fig:elastic_curvature}
\end{figure}

The velocity $v$ is obtained directly from the solution of the
equation of motion. To determine $r$, we first search for the correct circle
which touches the path and has the correct curvature as shown in
Fig.~\ref{fig:elastic_curvature}.

To identify the required circle we first write it in the form
\begin{subequations}
  \begin{equation}
    z(x) = \pm \sqrt {r^2 - (x-x_0)^2 } 
  \end{equation}
  and calculate its first and second derivatives,  
  \begin{align}
    \frac{\mathrm{d} z}{\mathrm{d} x}
    &= \mp \frac{x-x_0}{\sqrt {r^2 - (x-x_0)^2 }}
      = - \frac{x-x_0}{z-z_0} \; , \\
    \frac{\mathrm{d}^2 z}{\mathrm{d} x^2}
    &= - \frac{1}{z-z_0} \left ( 1
      + \left ( \frac{\mathrm{d} z}{\mathrm{d} x} \right )^2 \right ) \; .
  \end{align}
\end{subequations}
To describe the path locally in a correct way, the three equations
\begin{subequations}
  \begin{align}
    z(x) &= \pm \sqrt {r^2 - (x-x_0)^2 } = h(x) \; , \\
    \frac{\mathrm{d} z}{\mathrm{d} x} &= - \frac{x-x_0}{z-z_0} = h'(x) \; , \\
    \frac{\mathrm{d}^2 z}{\mathrm{d} x} &= - \frac{1}{z-z_0} \left ( 1 + h'(x)^2
                          \right ) = h''(x)
  \end{align}
\end{subequations}
have to be fulfilled. From these equations we obtain the radius of the circle
 \begin{equation}
   r = \frac{1+h'^2}{h''} \sqrt{1+h'^2} \; .
 \end{equation}
With the square of the velocity
\begin{equation}
  v^2 = \dot{x}^2 + \dot{z}^2 =  (1+h'^2) \dot{x}^2
\end{equation}
the radial acceleration reads
\begin{equation}
  a_r = \frac{h''}{1+h'^2}\sqrt{1+h'^2}\, \dot{x}^2 \; .
\end{equation}

In a last step we have to determine the directions of the accelerations to
extract the contribution perpendicular to the path. To do so, we introduce
the tangent vector (cf., Fig.~\ref{fig:elastic_curvature}),
\begin{subequations}
  \begin{align}
    \vec{t} &= \frac{1}{\sqrt{\mathrm{d} x^2 + \mathrm{d} z^2}} \begin{pmatrix} \mathrm{d} x \\
      \mathrm{d} z \end{pmatrix} =  \frac{1}{\sqrt{1 + h'(x)^2}} \begin{pmatrix} 1 \\
      h'(x)  \end{pmatrix}
  \end{align}
  and the normal vector
  \begin{equation}
    \vec{n} = \frac{1}{\sqrt{1 + h'(x)^2}} \begin{pmatrix} -h'(x) \\
      1 \end{pmatrix} \; .
  \end{equation}
\end{subequations}
Then the vectorial representation of the radial acceleration
\eqref{eq:radial_acc} reads
\begin{equation}
  \vec{a}_r = \frac{h''}{1+h'^2}\sqrt{1+h'^2}\,  \dot{x}^2 \, \vec{n}
  = \frac{h''}{1+h'^2} \dot{x}^2 \begin{pmatrix} -h'(x) \\ 1 \end{pmatrix}
\end{equation}
with the components
\begin{subequations}
  \begin{align}
    a_{r,x} &= -\frac{h'' h'}{1+h'^2}\, \dot{x}^2 \; , \\
    a_{r,z} &= \frac{h''}{1+h'^2}\, \dot{x}^2 \; ,
  \end{align}
\end{subequations}
and the normal component compensating the gravitational acceleration is
\begin{equation}
  g_n = g\vec{e}_z\cdot \vec{n} = \frac{g}{\sqrt{1+h'^2}} \; .
\end{equation}

\section{Potential of the extended train of fixed length}
\label{sec:app_fixed_length}

To enable quantitative calculations for the model introduced in
Sec.~III it is necessary to know the potential of the train when different
parts of it are on different positions on the track. To do so, we first assume
that the height can be determined as a function of $x$,
\begin{equation}
  z = h(x) \; ,
  \label{eq:zhx}
\end{equation}
and that the fixed-length train is defined by the position of its front
$x_\mathrm{F}$ and the length $L$, i.e.\ the position of the train end
$x_\mathrm{E}$  can be calculated via the integral
\begin{equation}
  \int_{x_\mathrm{E}}^{x_\mathrm{F}}\mathrm{d} s = \int_{x_\mathrm{E}}^{x_\mathrm{F}}
  \sqrt{1+h'(x)^2}\,\mathrm{d} x = F(x_\mathrm{F}) -F(x_\mathrm{E}) = L \;.
  \label{eq:fixed_length_int}
\end{equation}
Generally, this has to be done numerically, but formally we may write
\begin{equation}
  x_\mathrm{E} = F^{-1}\,\left (  F(x_\mathrm{F}) - L \right ) \; ,
  \label{eq:end_of_train}
\end{equation}
where $F(x)$ is the antiderivative of $\sqrt{1 + h'(x)^2}$ and $F^{-1}$ is
its inverse, respectively.

We assume the train has a homogeneous mass distribution with density $\mu$ and
total mass $M=\mu L$. Then the train's potential energy is determined via
\begin{equation}
  U(x_\mathrm{F}) = g\,\int_{x_\mathrm{E}}^{x_\mathrm{F}} \mu\,z\,\mathrm{d} s
  = g\,\mu\,\int_{x_\mathrm{E}}^{x_\mathrm{F}} h(x)\,
  \sqrt{1 + h'(x)^2}\,\mathrm{d} x \;.
  \label{eq:fixed_length_pot}
\end{equation}

The same calculation can be done for a parameterized curve,
\begin{equation}
  x = x(u)\; , \quad
  x'(u) = \frac{\mathrm{d} x}{\mathrm{d} u} \; , \quad
  z= z(u) \;, \quad
  z'(u) = \frac{\mathrm{d} z}{\mathrm{d} u}\; , \quad
  L =  \int_{u_\mathrm{E}}^{u_\mathrm{F}}\mathrm{d} s
  =  \int_{u_\mathrm{E}}^{u_\mathrm{F}} \sqrt{(x')^2 + (z')^2}\, \mathrm{d} u \; .
  \label{eq:fixed_length_parameters}
\end{equation}
Again we use the front of the train $u_\mathrm{F}$ as independent variable
and write
\begin{equation}
  u_\mathrm{E} = F^{-1}\left ( F(u_\mathrm{F})-L\right )
  \label{eq:fixed_length_uE}
\end{equation}
with
\begin{equation*}
  F(u) = \int \sqrt{x'(u)^2 + z'(u)^2}\,\mathrm{d} u \; .
\end{equation*}
The potential energy in this case is calculated with
\begin{equation}
  U = \mu\,g\,\int\limits_{u_\mathrm{E}}^{u_\mathrm{F}} z(u)\,
  \sqrt{x'(u)^2 + z'(u)^2} \,\mathrm{d} u \;.
  \label{eq:fixed_length_param_pot}
\end{equation}

The line integral in Eq.~\eqref{eq:fixed_length_param_pot} required to
calculate the potential energy indicates that examples for analytical
expressions for the potential are rare. However, for some track shapes
they can be obtained with short calculations. An especially appealing
example is a train of length $L$ on a circular track, for which we provide
the simple example of the potential. A good choice for the parameter $u$ is
the angle $\varphi_\mathrm{F}$ defining the front of the train,
\begin{equation}
  x_\mathrm{F} = R \sin(\varphi_\mathrm{F}) \; ,
  \qquad z_\mathrm{F} = R \left ( 1- \cos(\varphi_\mathrm{F}) \right ) \; .
  \label{eq:fixed_parametrization_circle}
\end{equation}
Depending on the position of $\varphi_\mathrm{F}$ the potential is
\begin{equation}
  U(\varphi_\mathrm{F}) =  \mu\,g\,R^2\, \left [ \frac{L}{R}
    - \sin\varphi_\mathrm{F} + \sin\left ( \varphi_\mathrm{F}
      - \frac{L}{R} \right ) \right ] \;.
  \label{eq:fixed_parametrization_pot}
\end{equation}
With this potential we can derive the equations of motion exactly the same
way as it is done for a point particle.

\section{Additional considerations for parameterized curves}

\subsection{Roller coaster as point particle}
\label{sec:point_particle_Loop}

In this section we derive the equations of motion for a point particle on
a track defined by a parameterized curve. This is, e.g., necessary if one
of the coordinates $x$ or $z$ does not unambiguously define the wagon's
position, e.g., for a loop, which is a very common curve for a roller coaster.

We start with the definition of our variables as functions of a parameter $u$,
\begin{align*}
  x = x(u) \quad \text{and} \quad z = z(u) \; .
\end{align*}
Derivatives lead us to the velocity and the acceleration,
\begin{subequations}
  \begin{align}
    v_x &= \frac{\partial x}{\partial u}\,\dot{u} = x'\;\dot{u} \; , \\
    v_z &= \frac{\partial z}{\partial u}\,\dot{u} = z'\;\dot{u} \;, \\
    a_x &= \frac{\partial^2 x}{\partial u^2}\,\dot{u}^2 + \frac{\partial x}
          {\partial u}\,\ddot{u}
          = x''\,\dot{u}^2 + x'\,\ddot{u}  \; , \\
    a_z &= \frac{\partial^2 z}{\partial u^2}\,\dot{u}^2 + \frac{\partial z}{\partial u}\,
          \ddot{u} = z''\,\dot{u}^2 + z'\,\ddot{u} \;. 
  \end{align} \label{eq:derivatives_parametric} 
\end{subequations}

A way of deriving the equation of motion for the parameter $u$ alternative to
the Lagrange formalism is the application of Newton's laws of motion. To
proceed in this direction we write the equations of motion in the coordinates
$x$ and $z$,
\begin{subequations}
  \begin{align}
    a_x &= -\frac{F_{N}}{m}\,\sin\theta \; , \\
    a_z &= \frac{F_{N}}{m}\,\cos\theta -g \; ,
          \label{eq:accelerations_parametric} 
  \end{align}
\end{subequations}
and remove the normal force $F_{N}$, which leads to an equation of motion
for the parameter $u$,
\begin{align}
  \ddot{u} = -\frac{\dot{u}^2\,(x'\,x'' + z'\,z'') - g\,z'}
  {(x')^2 + (z')^2} \;.
  \label{eq:eqs_motion_parametric} 
\end{align}
Even though it seems to be more complex, Eq.~\eqref{eq:eqs_motion_parametric}
is the equivalent to the Lagrangian equations of motion
(5a) and (5b) for a point particle on an arbitrary track that can be described
by the parameter $u$.

\begin{figure}[tbp]
  \includegraphics[width=0.5\columnwidth]{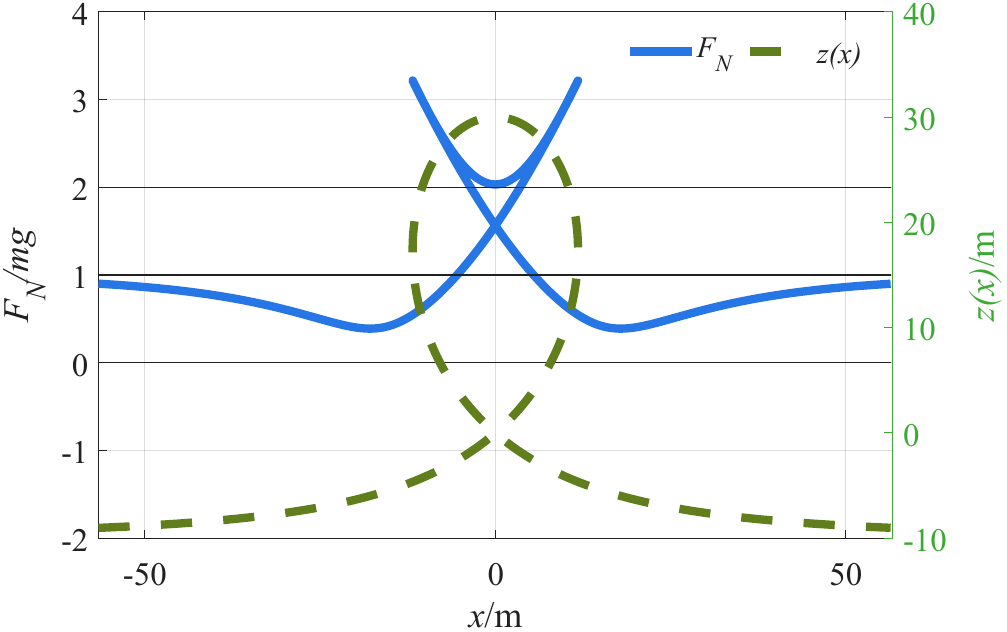}
  \caption{Roller coaster as a point particle on a MacLaurin trisectrix as
    an example for a train in a loop. The thin horizontal lines mark the
    values of $F_N/mg = 0$, $F_N/mg = 1$ and $F_N/mg=2$.}
  \label{fig:trixectrix}
\end{figure}

A very typical example is the MacLaurin trisectrix, which is given by
\begin{align}
  x(u) &= \frac{b\,u\,\left ( u^2-3\right )}{u^2+1} \; ,  \\
  z(u) &= -\frac{b\,\left ( u^2-3\right )}{u^2+1} \; ,
\end{align}
where $b$ approximately describes the height of the trisectrix, and $u$ is
the parameter of the track. The derivatives required in
Eq.~\eqref{eq:eqs_motion_parametric} are,
\begin{subequations}
    \begin{align}
      x' &= \frac{b\,\left ( u^4 + 6u^2 -3\right )}{\left ( u^2+1\right )^2}
           \;, \\
      x'' &= \frac{6\,b\,u}{u^2+1}\,\left ( 1- \frac{\frac{7}{3}\,u^2-3}{u^2+1}
            + \frac{\frac{4}{3}\,u^2\,\left ( u^2-3\right )}
            {\left ( u^2+1\right )^2}\right )\;,  \\
      z' &= \frac{8\,b\,u}{\left ( u^2 +1\right )^2} \;, \\
      z'' &= \frac{2b}{u^2+1}\,\left (  1-\frac{5u^2-3}{u^2+1}
            + \frac{4u^2\,\left (  u^2-3\right )}
            {\left (  u^2+1\right )^2}\right ) \;. 
    \end{align}
\end{subequations}
With these terms the equation of motion \eqref{eq:eqs_motion_parametric} can
be integrated numerically. From \eqref{eq:derivatives_parametric} and
\eqref{eq:accelerations_parametric} the velocities, accelerations and normal
forces on the trisectrix loop can be determined. An example for the normal
forces experienced in the trisectrix is given in
Fig.~\ref{fig:trixectrix}.

\subsection{Wagons connected via springs}
\label{sec:app_springs_parametric}

For the circular shaped track mentioned in Sec.~\ref{sec:app_fixed_length}
we need to express the equations of motion for the model consisting of
several wagons connected via springs in terms of a parameter $u$, i.e.,
we have $x(u)$ and $z(u)$.

First we write the equations of motion \eqref{eq:springs_eqsm_x1}
and \eqref{eq:springs_eqsm_z1} for wagon $i=1$ with a general form of a normal
force $\vec{F}_N$, 
\begin{subequations}
  \begin{align}
    \ddot{x}_1 &= \frac{k}{m}\,\frac{\sqrt{\left ( x_2-x_1\right )^2
                 + \left ( z_2-z_1\right )^2} -s_0}{\sqrt{\left (
                 x_2-x_1\right )^2 + \left ( z_2-z_1\right )^2}}\,
                 \left ( x_2-x_1\right ) - \frac{F_N}{m} \sin\vartheta_1\;,
                 \label{eq:springs_eqsm_gx1} \\
    \ddot{z}_1 &= \frac{k}{m}\,\frac{\sqrt{\left ( x_2-x_1\right )^2
                 + \left ( z_2-z_1\right )^2} -s_0}{\sqrt{\left (
                 x_2-x_1\right )^2 + \left ( z_2-z_1\right )^2}}
                 \left ( z_2 - z_1\right ) -g + \frac{F_N}{m} \cos\vartheta_1
                 \;. \label{eq:springs_eqsm_gz1}
  \end{align}
  We call the angle between the tangent to the track at point $(x_1,z_1)$ and
  the $x$ axis $\vartheta_1$. The normal force is perpendicular
  to the tangent. In a similar way we find for the last wagon ($i=N$)
  \begin{align}
    m\,\ddot{x}_N &= -\frac{k}{m} \frac{\sqrt{\left ( x_N - x_{N-1}\right )^2
                    + \left ( z_N - z_{N-1}\right )^2} - s_0}
                    {\sqrt{\left ( x_N - x_{N-1}\right )^2
                    + \left ( z_N - z_{N-1}\right )^2}}
                    \left ( x_N - x_{N-1}\right )
                    - \frac{F_N}{m} \sin\vartheta_N \;, \\
    m\,\ddot{z}_N &= -\frac{k}{m} \frac{\sqrt{\left ( x_N - x_{N-1}\right )^2
                    + \left ( z_N - z_{N-1}\right )^2} - s_0}{\sqrt{\left(
                    x_N - x_{N-1}\right )^2 + \left ( z_N - z_{N-1}\right )^2}}
                   \left ( z_N - z_{N-1}\right ) - m\,g
                    + \frac{F_N}{m} \cos\vartheta_N \; ,
  \end{align}
  and for all middle wagons ($2\le i \le N-1$)
    \allowdisplaybreaks
    \begin{align}
      m\,\ddot{x}_i
      &= -\frac{k}{m}\, \frac{\sqrt{\left ( x_i - x_{i-1}
        \right )^2 + \left ( z_i- z_{i-1}\right )^2} - s_0}
        {\sqrt{\left ( x_i - x_{i-1}\right )^2 + \left ( z_i - z_{i-1}
        \right )^2}} \,\left ( x_i -x_{i-1}\right ) \notag \\
      &\quad + \frac{k}{m}\, \frac{\sqrt{\left ( x_{i+1} - x_i \right )^2
        + \left ( z_{i+1}- z_i \right )^2} - s_0}{\sqrt{\left ( x_{i+1}
        - x_i \right )^2 + \left ( z_{i+1} - z_i \right )^2}} \,
        \left ( x_{i+1}-x_i\right ) - \frac{F_N}{m} \sin\vartheta_i \;, \\
      m\,\ddot{z}_i
      &= -\frac{k}{m}\, \frac{\sqrt{\left ( x_i - x_{i-1}
        \right )^2 + \left ( z_i- z_{i-1}\right )^2} - s_0}{\sqrt{
        \left ( x_i - x_{i-1}\right )^2 + \left ( z_i - z_{i-1}\right )^2}}
        \,\left ( z_i - z_{i-1} \right ) \notag \\
      &\quad +\frac{k}{m}\, \frac{\sqrt{\left ( x_{i+1} - x_i \right )^2
        + \left ( z_{i+1}- z_i \right )^2} - s_0}{\sqrt{\left ( x_{i+1}
        - x_i \right )^2 + \left ( z_{i+1} - z_i \right )^2}}\,
        \left ( z_{i+1} - z_i\right ) -m\,g + \frac{F_N}{m} \cos\vartheta_N \;.
    \end{align}
\end{subequations}

The parametrization of the track is defined by
{\allowdisplaybreaks
  \begin{align*}
    x_i &= x_i(u_i)\;, & z_i &= z_i(u_i)\;, \\
    \dot{x}_i &= x_i'(u_i)\;\dot{u}_i\;, & \dot{z}_i
                                   &= z_i'(u_i)\;\dot{u}_i \;, \\
    \ddot{x}_i &= x_i''(u_i)\;\dot{u}_i^2 + x_i'(u)\;\ddot{u}_i\;,
                       & \ddot{z}_i &= z_i''(u_i)\;\dot{u}_i^2
                                            + z_i'(u_i)\;\ddot{u}_i \;. \\
  \end{align*}
}
The calculations to write the equations of motion in terms of the parameter
are very similar for all wagons $i$ and always follow the same principle.
As an example we do this calculation for the first wagon $i=1$, i.e., for the
equations of motion \eqref{eq:springs_eqsm_gx1} and
\eqref{eq:springs_eqsm_gz1}, which we want to express by the parameter
$u_1$. To do so, we first write for \eqref{eq:springs_eqsm_gx1}
\begin{align*}
  \ddot{x}_1 &= x_1''(u_1)\dot{u}_1^2 + x_1'(u)\ddot{u}_1 \\
             &= \frac{k}{m}\,\frac{\sqrt{\left ( x_2-x_1\right )^2
               + \left ( z_2-z_1\right )^2} -s_0}{\sqrt{
               \left ( x_2-x_1\right )^2 + \left ( z_2-z_1\right )^2}}\,
               \left ( x_2-x_1\right ) \\
             &\quad- \frac{F_N}{m} \cos\vartheta_1 \tan\vartheta_1 \; .
\end{align*}
The term of the normal force is written in a way, in which we can
identify the identical term $(F_1/m)\cos\vartheta_1$ in
Eq.~\eqref{eq:springs_eqsm_gz1}. We use this identity to replace the
unknown normal force from our equation and exploit that
\begin{equation*}
  \tan\vartheta = \frac{z'}{x'}\; ,
\end{equation*}
which leads to
\begin{multline*}
  \ddot{u}_1 = \Bigg [ -\left(x_1'\,x_1''+ z_1'\,z_1'' \right ) \dot{u}_1^2
  + \frac{k}{m}\,\frac{\sqrt{\left ( x_2-x_1\right )^2 + \left ( z_2-z_1\right )^2} -s_0}{\sqrt{\left ( x_2-x_1\right )^2 + \left ( z_2-z_1\right )^2}}\,\left ( x_2-x_1\right ) x_1' \\
  + \frac{k}{m}\,\frac{\sqrt{\left ( x_2-x_1\right )^2 + \left ( z_2-z_1\right )^2} -s_0}{\sqrt{\left ( x_2-x_1\right )^2 + \left ( z_2-z_1\right )^2}}\,\left ( z_2-z_1\right ) z_1' - g z_1' \Bigg ]
/ \left ( x_1'^2 + z_1'^2 \right ) \; .
\end{multline*}

Analogously we calculate in the case $i=N$
\begin{multline*}
  \ddot{u}_N = \Bigg [ -\left(x_N'\,x_N''+ z_N'\,z_N'' \right ) \dot{u}_N^2
  - \frac{k}{m}\,\frac{\sqrt{\left ( x_N-x_{N-1}\right )^2
      + \left ( z_N-z_{N-1}\right )^2} -s_0}{\sqrt{\left ( x_N-x_{N-1}
      \right )^2 + \left ( z_N-z_{N-1} \right )^2}}\,\left ( x_N-x_{N-1}
  \right ) x_N' \\
  - \frac{k}{m}\,\frac{\sqrt{\left ( x_N-x_{N-1}\right )^2
      + \left ( z_N-z_{N-1}\right )^2} -s_0}{\sqrt{\left ( x_N-x_{N-1}
      \right )^2 + \left ( z_N-z_{N-1} \right )^2}}\,\left ( z_N-z_{N-1}
  \right ) z_N' - g z_N' \Bigg ]  / \left ( x_N'^2 + z_N'^2 \right )
\end{multline*}
and in all other cases ($2\le i \le  N-1$)
\begin{multline*}
  \ddot{u}_i = \Bigg [ -\left(x_i'\,x_i''+ z_i'\,z_i'' \right ) \dot{u}_i^2 
  - \frac{k}{m}\,\frac{\sqrt{\left ( x_i-x_{i-1}\right )^2
      + \left ( z_i-z_{i-1}\right )^2} -s_0}{\sqrt{\left ( x_i-x_{i-1}
      \right )^2 + \left ( z_i-z_{i-1} \right )^2}}\,\left ( x_i-x_{i-1}
  \right ) x_i' \\
  + \frac{k}{m}\,\frac{\sqrt{\left ( x_{i+1}-x_i\right )^2
      + \left ( z_{i+1}-z_i\right )^2} -s_0}{\sqrt{\left ( x_{i+1}-x_i
      \right )^2 + \left ( z_{i+1}-z_i \right )^2}}\,\left ( x_{i+1}-x_i
  \right ) x_i' 
  - \frac{k}{m}\,\frac{\sqrt{\left ( x_i-x_{i-1}\right )^2
      + \left ( z_i-z_{i-1}\right )^2} -s_0}{\sqrt{\left ( x_i-x_{i-1}
      \right )^2 + \left ( z_i-z_{i-1} \right )^2}}\,\left ( z_i-z_{i-1}
  \right ) z_i' \\
  + \frac{k}{m}\,\frac{\sqrt{\left ( x_{i+1}-x_i\right )^2
      + \left ( z_{i+1}-z_i\right )^2} -s_0}{\sqrt{\left ( x_{i+1}-x_i
      \right )^2 + \left ( z_{i+1}-z_i \right )^2}}\,\left ( z_{i+1}-z_i
  \right ) z_i' - g z_i' \Bigg ] / \left ( x_i'^2 + z_i'^2 \right ) \; .
\end{multline*}

\begin{figure}[tbp]
  \centering
  \includegraphics[width=0.5\columnwidth]{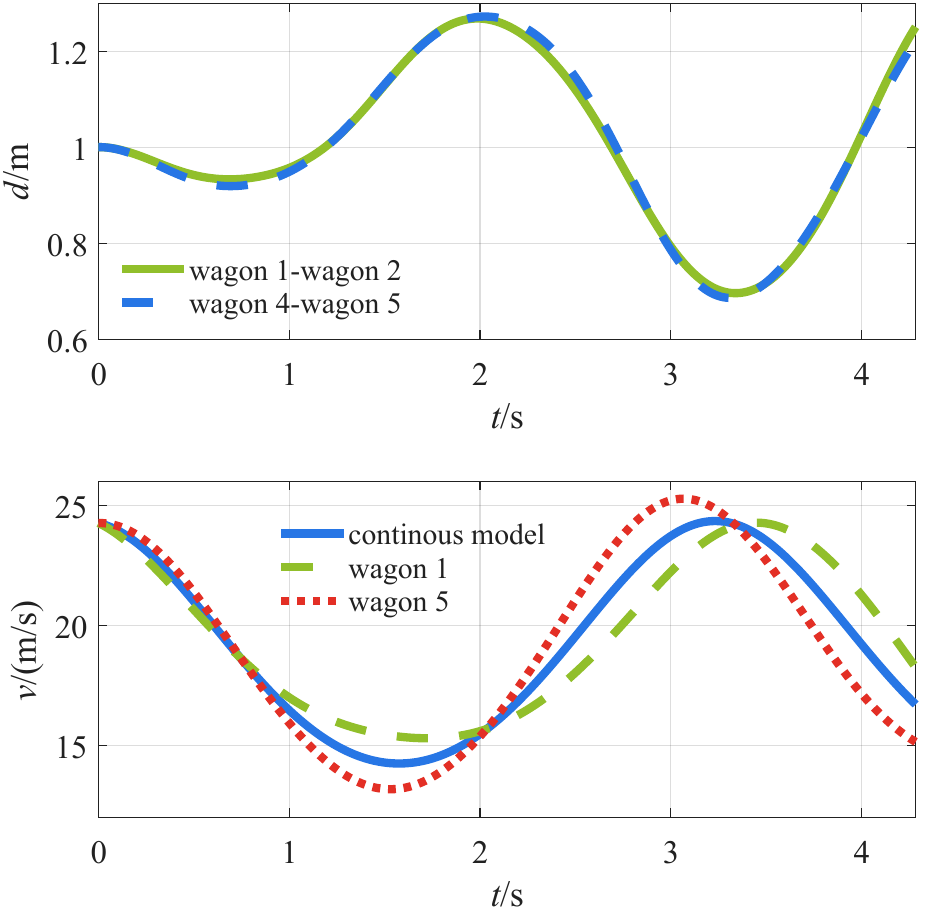}
  \caption{Motion of a roller coaster train in a circular loop, where the
    parameters are chosen such that the springs of between the wagons are
    stressed noticeably. Deviations from the fixed-length model are visible,
    however the overall motion is the same. The circle has a radius of
    $\SI{10}{\metre}$. The train consists of 5 wagons with an equilibrium
    distance of $s_0 = \SI{1}{\metre}$ and the spring constant for a spring
    between two wagons is $k=\SI{10}{\kilo\newton \per \metre}$.}
  \label{fig:comparison_2models1}
\end{figure}

With these equations of motion we are able to compare the model to that
of a fixed-length train with a numerical solution of the problem.
In the discrete wagons model we chose a train consisting of 5 wagons with the
same parameters as in Sec.~IV. The result is shown in
Fig.~\ref{fig:comparison_2models1}. 
With the chosen spring constant, relative motions (oscillations) of the wagons
are possible as can be seen in Fig.~\ref{fig:comparison_2models1}a. Of course,
this effect is not present in the fixed-length model. This can be observed in
Fig.~\ref{fig:comparison_2models1}b, in which the velocities of selected
wagons is shown as a function of time. None of the wagons has exactly the same
progression of its velocity as that of the fixed-length train. However, the
overall agreement between the models is very good. The single wagons are very
close to the motion of the fixed-length train.

We checked for a spring constant which is artificially set to a high
value such that the springs are hardly elongated that both models agree even
better. Indeed, our numerical calculations showed that the model with
springs seem clearly to converge to the fixed-length model if the springs
approach higher and higher values.

\end{document}